\documentclass{article} %
\usepackage{iclr2027_conference,times}
\iclrfinalcopy

\usepackage{amsmath,amsfonts,bm}

\def\eqref#1{equation~\ref{#1}}

\def\1{\bm{1}}

\def\vzero{{\bm{0}}}

\def\vb{{\bm{b}}}
\def\vc{{\bm{c}}}

\def\vn{{\bm{n}}}

\def\vp{{\bm{p}}}
\def\vq{{\bm{q}}}

\def\vu{{\bm{u}}}
\def\vv{{\bm{v}}}
\def\vw{{\bm{w}}}
\def\vx{{\bm{x}}}
\def\vy{{\bm{y}}}
\def\vz{{\bm{z}}}

\DeclareMathAlphabet{\mathsfit}{\encodingdefault}{\sfdefault}{m}{sl}
\SetMathAlphabet{\mathsfit}{bold}{\encodingdefault}{\sfdefault}{bx}{n}

\def\gB{{\mathcal{B}}}
\def\gC{{\mathcal{C}}}

\def\gH{{\mathcal{H}}}

\def\gP{{\mathcal{P}}}

\def\gS{{\mathcal{S}}}

\def\gW{{\mathcal{W}}}
\def\gX{{\mathcal{X}}}
\def\gY{{\mathcal{Y}}}
\def\gZ{{\mathcal{Z}}}

\newcommand{\R}{\mathbb{R}}

\DeclareMathOperator*{\argmin}{arg\,min}

\usepackage{hyperref}
\usepackage{url}

\title{Learning Variable Metrics for Convergent Proximal Splitting Algorithms}

\author{Yodai Suzuki \& Shunsuke Ono\\
Department of Computer Science\\
Institute of Science Tokyo\\
4259 Nagatsuta-cho, Midori-ku, Yokohama, Kanagawa 226-8501, Japan \\
\texttt{\{suzuki.y.bd8c, ono.s.5af2\}@m.isct.ac.jp}
}

\usepackage{amsthm}
\usepackage{amssymb}
\usepackage{booktabs}
\usepackage{tabularx}
\usepackage{graphicx}
\usepackage{tikz}
\usetikzlibrary{arrows.meta}

\newtheorem{theorem}{Theorem}[section]
\newtheorem{proposition}[theorem]{Proposition}
\newtheorem{lemma}[theorem]{Lemma}
\theoremstyle{definition}
\newtheorem{remark}[theorem]{Remark}

\newcommand{\Id}{\mathrm{Id}}
\DeclareMathOperator{\prox}{prox}
\DeclareMathOperator{\zer}{zer}
\DeclareMathOperator{\gra}{gra}
\DeclareMathOperator{\Diag}{Diag}

\begin{document}

\maketitle
\lhead{}

\begin{abstract}
We propose learned variable metric (LVM), a framework that learns metrics of proximal splitting algorithms adapted to each problem instance, coordinate, and iteration with a convergence guarantee.
Proximal splitting algorithms can solve a wide variety of convex optimization problems but often require many iterations to converge.
Choosing a suitable metric is an effective remedy, and variable metric theory naturally extends convergence guarantees to metrics that vary over the iterations.
Learning is a promising way to design such adaptive metrics beyond the reach of manual design, but a naive implementation leads to (i) an expensive evaluation of a network at every iteration and (ii) difficulty in satisfying the conditions that the theory imposes on the whole infinite sequence of metrics. To resolve these difficulties, we first train a neural network to generate a few representative metrics once per problem instance, and then obtain the metric of every iteration by combining them with basis functions of the normalized iteration index.
We present LVM for the proximal gradient method with a quadratic term and the primal-dual hybrid gradient method, and prove their convergence under a mild spectral condition, which is ensured by a rescaling of the representative metrics.
In experiments on deblurring and inpainting, LVM reduces the number of iterations to reach a target accuracy in most tested settings compared with several baselines, and the computation time at high accuracies.
\end{abstract}

\section{Introduction}
\label{sec:intro}

A vast range of problems in signal processing, imaging, and data science are
formulated as the convex minimization problem
\begin{equation}
\label{eq:problem}
  \underset{\vx\in\gX}{\text{minimize}}\quad f(\vx)+g(L\vx),
\end{equation}
where $f$ and $g$ are proper lower-semicontinuous convex functions on real
Hilbert spaces $\gX$ and $\gY$, respectively, and $L\colon\gX\to\gY$ is a bounded linear
operator. This formulation has a wide variety of applications, e.g., inverse problems in
imaging such as deblurring, inpainting, compressed sensing, and tomographic
reconstruction
\citep{Afonso2011,Komodakis2015_review,Chambolle_Pock_2016_review}, sparse
coding \citep{BeckTeboulle2009,Veshki2022}, and hyperspectral image restoration~\citep{Naganuma2023} and unmixing~\citep{Iordache2012, Naganuma2024HSUnmixing}.
Since~(\ref{eq:problem}) typically contains nondifferentiable terms, proximal splitting algorithms,
which handle such terms through their proximity operators, have been studied extensively under diverse assumptions on $f$, $g$, and
$L$~\citep{Condat2013,Vu2013PDS,Combettes2024rev}.
A typical setting assumes that $L=\Id$ and $f$ is differentiable
with a Lipschitz continuous gradient, in which case the proximal gradient (PG)
method is available \citep{LionsMercier1979,Chen1997,Bauschke2017}.
For a general $L$, the primal-dual hybrid gradient (PDHG) method is widely
used, which avoids an inversion induced by $L$ at each iteration~\citep{ChambollePock2011}.

Despite their flexibility, proximal splitting algorithms often require a large number of iterations to obtain an accurate solution \citep{Pock2011Prec,Themelis2019,Zhang2020}.
Accelerating their convergence has therefore received considerable attention, as we review below.

\begin{table}[t]
  \centering
  \fontsize{8}{9.5}\selectfont
  \caption{Properties of representative acceleration schemes and ours. Parentheses indicate a dependency of the property on the work.}
  \vspace{1mm}
  \label{tab:comparison}
  \setlength{\tabcolsep}{2pt}
  \renewcommand{\arraystretch}{1}
  \begin{tabularx}{\linewidth}{>{\raggedright\arraybackslash}X c c c c c}
    \toprule
    \textbf{Scheme}
    & \begin{tabular}[b]{@{}c@{}}\textbf{Problem-}\\\textbf{specific}$^{\dagger}$\end{tabular}
    & \begin{tabular}[b]{@{}c@{}}\textbf{Coordinate-}\\\textbf{wise}\end{tabular}
    & \begin{tabular}[b]{@{}c@{}}\textbf{Iteration-}\\\textbf{varying}\end{tabular}
    & \begin{tabular}[b]{@{}c@{}}\textbf{No extra}\\\textbf{evaluations}$^{\ddagger}$\end{tabular}
    & \begin{tabular}[b]{@{}c@{}}\textbf{Convergence}\\\textbf{guarantee}\end{tabular} \\
    \midrule
    Inertial acceleration~\citep[e.g.,][]{BeckTeboulle2009}
      & & & $\checkmark$ & $\checkmark$ & $\checkmark$ \\
    Preconditioning~\citep[e.g.,][]{Pock2011Prec}
      & $\checkmark$ & $\checkmark$ & & $\checkmark$ & $\checkmark$ \\
    \midrule
    Line search~\citep[e.g.,][]{Malitsky2018}
      & $\checkmark$ & & $\checkmark$ & & $\checkmark$ \\
    Residual balancing~\citep{Goldstein2015}
      & $\checkmark$ & & $\checkmark$ & $\checkmark$ & $\checkmark$ \\
    Curvature estimation~\citep{Malitsky2024}
      & $\checkmark$ & & $\checkmark$ & $\checkmark$ & $\checkmark$ \\
    \midrule
    Learned maps~\citep[e.g.,][]{Banert2024}
      & $\checkmark$ & $\checkmark$ & $\checkmark$ & & $\checkmark$ \\
    Learned stepsizes~\citep[e.g.,][]{Sambharya2026}
      & & & $\checkmark$ & $\checkmark$ & ($\checkmark$) \\
    Learned fixed metrics~\citep[e.g.,][]{King2024}
      & $\checkmark$ & $\checkmark$ & & $\checkmark$ & $\checkmark$ \\
    \midrule
    \textbf{LVM (ours)}
      & $\boldsymbol{\checkmark}$ & $\boldsymbol{\checkmark}$ & $\boldsymbol{\checkmark}$ & $\boldsymbol{\checkmark}$ & $\boldsymbol{\checkmark}$ \\
    \bottomrule
  \end{tabularx}
  \par\vspace{1mm}
  {\raggedright\footnotesize
  $^{\dagger}$Adapted to each problem instance.\quad
  $^{\ddagger}$No additional evaluation of operators or networks per iteration.\par}
  \vspace{-3mm}
\end{table}

\subsection{Related Work}
\label{sec:related}
\paragraph{Acceleration for Proximal Splitting Algorithms.}
Table~\ref{tab:comparison} summarizes representative schemes.
One of the most well-established approaches for acceleration introduces an inertial term that does not depend on the problem structure, as in Nesterov's acceleration~\citep{BeckTeboulle2009,Parikh2014_sig,Donoghue2015} and the heavy-ball method~\citep{Alvarez2001, Bot2015}.
Preconditioning also speeds up convergence~\citep{Liu2021prec,Bredies2022} by replacing the metric of the underlying space with one adapted to the problem, e.g., to set stepsizes coordinate-wise \citep{Pock2011Prec} or to exploit the block structure of $L$ \citep{Wen2017prec,Naganuma2023}.
Although preconditioning applies to a wide range of algorithms, its metric is determined by hand-designed rules and does not change along the iterations.
Another line of work adapts the stepsizes to the current iterate, to take larger steps by estimating the local Lipschitz constant with line search~\citep{Malitsky2018,Chang2022}.
However, these schemes may require repeated trial updates at every iteration, which can be expensive in practice.
Other schemes balance the multiple stepsizes in PDHG~\citep{Goldstein2015} or estimate the local Lipschitz constant from consecutive iterates~\citep{Malitsky2020,Malitsky2024}, but they adapt only scalar stepsizes.

\paragraph{Learned Solvers.}
The success of deep learning in recent years opens a new direction for solving~(\ref{eq:problem}), as the expressive power of neural networks allows complex rules of algorithms to be learned from data.
The dominant approach is unrolling, which truncates the algorithm to a fixed number of iterations and trains the parameters end to end \citep{Gregor2010LISTA,Adler2017LearnedPR,Cherkaoui2020, Monga2021Unrolling}.
It is typically trained so that the output well approximates a solution to~(\ref{eq:problem}) or directly approximates the desired solution modeled by~(\ref{eq:problem})~\citep{Xiaohan2018}, but asymptotic convergence beyond the trained depth falls outside its standard formulation.
Another attempt learns a map appearing in the update, e.g., a descent direction, and evaluates the network at every iteration to accelerate the algorithm.
These methods ensure convergence by, e.g., confining the output to a certain set \citep{Banert2024}, by a safeguard that falls back on an original update when a criterion fails \citep{Heaton2023}, or by restricting the learned geometry~\citep{Tan2023}.
Since such an evaluation of a network at every iteration typically costs much more than one iteration of the algorithm, several works instead learn a smaller part of the algorithm.
The stepsizes are one such target, adaptive over the iterations but shared by all the coordinates and the problem instances \citep{Ablin2019, Sambharya2026}, and a similar idea applies to the coefficients of a generalized primal-dual scheme \citep{Banert2020}.
Other works learn a metric that adjusts each coordinate but is fixed over the iterations~(\citealp[for linear solvers]{Li2023ICML}; \citealp{King2024}).

\paragraph{Variable Metric Algorithms.}
Variable metric theory studies algorithms whose metric changes along the iterations \citep{Bonnans1995VM,Combettes2012VM,Combettes2014ICIP}.
One of its central topics is their convergence~\citep{Combettes2014VMFBS, Suzuki2026VMPDS}, which can be established under several assumptions, including the requirement that the metrics decrease only by summable amounts between consecutive iterations (detailed in Section~\ref{sec:vm}).
Under this framework, several works design such adaptive metrics for acceleration with convergence guarantees, by a quadratic majorant of the smooth term~\citep{Chouzenoux2014}, a line search~\citep{Bonettini2016,Salzo2017}, or a quasi-Newton update~\citep{Becker2019}.
When metrics are restricted to scalar ones or diagonal ones with a block structure, the theory also supplies the analysis behind the adaptive stepsize rules, leading to, e.g., a convergent, spectrally motivated tuning for primal-dual Douglas--Rachford splitting~\citep{Lorenz2025} and a convergent stochastic line-search strategy for PDHG~\citep{Suzuki2026VMPDS}.

\subsection{Contributions}
\label{sec:contributions}

As the above discussion suggests, learning the metrics may provide a way to achieve complex adaptation to each problem instance, coordinate, and iteration beyond the reach of manual design, potentially with a convergence guarantee from variable metric theory.
However, a naive implementation of this approach faces two difficulties.
First, if the metrics are produced by network evaluations during the iterations, the acceleration can be impractical since evaluating a network typically costs much more than a single iteration of these algorithms.
Second, variable metric theory imposes conditions on the whole infinite sequence of metrics, which are difficult to ensure by training alone.

These difficulties motivate the central question of this paper. \emph{How can we learn problem-specific, coordinate-wise, and iteration-varying metrics of proximal splitting algorithms without extra evaluations per iteration and with a convergence guarantee?}
To answer this question, we propose a framework called learned variable metric (LVM).
Our contributions are summarized as follows.

\begin{itemize}
  \item
  We propose a method that generates a metric sequence varying over all the iterations by evaluating a neural network only once before the iterations.
  To this end, we design a few basis functions over the iterations, and introduce representative metrics for each basis function.
  The metric of every iteration is then obtained by combining the representative metrics with the basis functions.
  Each metric is also designed to be diagonal with a structure that keeps the proximity operators relative to it efficiently computable.
  These approaches make it possible to use the problem-specific, coordinate-wise, and iteration-varying metrics without incurring substantial additional cost at each iteration~(Section~\ref{sec:representative}).

  \item We develop specializations of LVM to PG with a quadratic term and PDHG. To satisfy the conditions for convergence, which reduce to those on the representative metrics by the construction of the sequence, we rescale them by an efficiently computable spectral factor. We also prove convergence of both algorithms when the rescaling is exact (Section~\ref{sec:convergence}).

  \item We train the network with a loss that measures the distance to a reference solution at several iterations of the algorithm.
  This loss is designed to accelerate convergence over a wide range of iterations by taking the logarithm of the distance~(Section~\ref{sec:training}).
\end{itemize}

Experiments on image deblurring and inpainting show that LVM needs fewer iterations than several baselines in most tested settings, and that its one-time setup cost pays off at high target accuracies.
To the best of our knowledge, LVM is the first framework that provides convergence guarantees for proximal splitting algorithms with learned coordinate-wise metrics that are adapted to each problem instance and vary over all the iterations without additional evaluations per iteration.

\section{Preliminaries}
\label{sec:prelim}
Throughout, $\gX$ and $\gY$ are real Hilbert spaces equipped with an inner
product $\langle\cdot\mid\cdot\rangle$ and its induced norm $\|\cdot\|$. We denote by
$\gB(\gX,\gY)$ the space of bounded linear operators from $\gX$ to $\gY$, by
$\gS(\gX)$ the set of self-adjoint operators in $\gB(\gX,\gX)$, and by $\Id$ the
identity operator.
Let $\Gamma_0(\gX)$ be the set of proper lower-semicontinuous convex functions from $\gX$ to $\R\cup\{+\infty\}$, and $\ell_{+}^{1}(\mathbb{N})$ be the set of summable sequences in $[0,+\infty)$. The
Fenchel--Rockafellar conjugate of $f\in\Gamma_0(\gX)$ is defined as $f^{*}(\vx):=\sup_{\vu\in\gX}(\langle\vx\mid\vu\rangle-f(\vu))$. The L\"owner partial ordering on $\gS(\gX)$ is 
\begin{equation}
\label{eq:loewner}
  U_1\succeq U_2\iff(\forall\vx\in\gX)\quad\langle U_1\vx\mid\vx\rangle\ge\langle U_2\vx\mid\vx\rangle,
\end{equation}
and we set $\gP_\alpha(\gX):=\{U\in\gS(\gX)\mid U\succeq\alpha\Id\}$ for
$\alpha>0$. Every $U\in\gP_\alpha(\gX)$ induces the inner product
$\langle\vx_1\mid\vx_2\rangle_{U}:=\langle\vx_1\mid U\vx_2\rangle$ and its
induced norm $\|\cdot\|_{U}$. The proximity operator of $f\in\Gamma_0(\gX)$
relative to the metric induced by $U$ is
\begin{equation}
\label{eq:prox}
  \prox_{f}^{U}(\vx):=\argmin_{\vu\in\gX}\ f(\vu)+\tfrac12\|\vx-\vu\|_{U}^{2},
\end{equation}
and we write $\prox_{f}:=\prox_{f}^{\Id}$. A function is called proximable when
its proximity operator can be computed efficiently. More
details are collected in Appendix~\ref{app:notation}.

\subsection{Proximal Splitting Algorithms on Variable Metrics}
\label{sec:vm}

In (\ref{eq:problem}), we assume that $f\in\Gamma_0(\gX)$, $g\in\Gamma_0(\gY)$,
$L\in\gB(\gX,\gY)$, and that its solution set is nonempty. Both PG and PDHG
admit variable metric extensions, as we state below.

\paragraph{PG on variable metrics with a quadratic term.}
Suppose in (\ref{eq:problem}) that $L=\Id$ and that
$f=\frac12\|A\cdot-\vb\|^{2}$, where $A\in\gB(\gX,\gW)$ and $\vb\in\gW$ for
a real Hilbert space $\gW$. With metrics $P_n\in\gP_\gamma(\gX)$, the updates are given as
\begin{equation}
\label{eq:vmpg}
  \vx_{n+1}=\prox_{g}^{P_n^{-1}}\left(\vx_n-P_nA^{*}(A\vx_n-\vb)\right).
\end{equation}
The following result states the convergence of (\ref{eq:vmpg}), and is proved in
Appendix~\ref{app:pg}.

\begin{theorem}
\label{thm:pg}
Let $\gamma>0$ and let $\zeta>0$. Suppose that the solution set to
(\ref{eq:problem}) is nonempty. Assume that the following hold:
\begin{enumerate}
  \item[(i)] $(\forall n\in\mathbb{N})$ $P_n\in\gP_\gamma(\gX)$.
  \item[(ii)] $(\exists\{\eta_n\}\in\ell_{+}^{1}(\mathbb{N}))(\forall n\in\mathbb{N})$
        $(1+\eta_n)P_{n+1}\succeq P_n$.
  \item[(iii)] $(\forall n\in\mathbb{N})$
        $P_n^{-1}-\frac12A^{*}A\succeq\zeta\Id$.
\end{enumerate}
Then $\{\vx_n\}$ defined by (\ref{eq:vmpg}) weakly converges to a solution of
(\ref{eq:problem}).
\end{theorem}

\paragraph{PDHG on variable metrics.}
Consider now the case where $f$ and $g$ in (\ref{eq:problem}) are both proximable
and $L$ is arbitrary. We use the primal-dual inclusion associated with
(\ref{eq:problem}),
\begin{equation}
\label{eq:inclusion}
  \text{find }(\hat{\vx},\hat{\vy})\in\gX\times\gY\ \text{ such that }\
  \begin{pmatrix}0\\0\end{pmatrix}\in
  \begin{pmatrix}\partial f(\hat{\vx})+L^{*}\hat{\vy}\\
                 -L\hat{\vx}+\partial g^{*}(\hat{\vy})\end{pmatrix},
\end{equation}
whose solution set consists of the associated primal-dual solution pairs under
mild qualification \citep{ChambollePock2011,Condat2013,Vu2013PDS}.
With metrics $P_n\in\gP_\gamma(\gX)$ and $Q_n\in\gP_\delta(\gY)$, the updates
are given as
\begin{equation}
\label{eq:vmpdhg}
\left\lfloor
\begin{array}{l}
  \vx_{n+1}=\prox_{f}^{P_n^{-1}}\left(\vx_n-P_nL^{*}\vy_n\right),\\[2pt]
  \vy_{n+1}=\prox_{g^{*}}^{Q_n^{-1}}\left(\vy_n+Q_nL(2\vx_{n+1}-\vx_n)\right).
\end{array}
\right.
\end{equation}
The following result states the convergence of (\ref{eq:vmpdhg}). It is the
convergence theorem of \citet{Suzuki2026VMPDS} with its assumption on
$\|P_n\|^{-1}\|Q_n\|^{-1}-\|L\|^{2}$ replaced by the spectral condition (iii) on
the block operator $V_n$, which is the form our framework satisfies directly.
The proof is discussed in Appendix~\ref{app:pdhg}.

\begin{theorem}
\label{thm:pdhg}
Let $\gamma>0$, let $\delta>0$, and let $\varsigma>0$. Suppose that the solution
set to (\ref{eq:inclusion}) is nonempty. Assume that the following hold:
\begin{enumerate}
  \item[(i)] $(\forall n\in\mathbb{N})$ $P_n\in\gP_\gamma(\gX)$,
        $Q_n\in\gP_\delta(\gY)$.
  \item[(ii)] $(\exists\{\eta_n^{(1)}\},\{\eta_n^{(2)}\}\in\ell_{+}^{1}(\mathbb{N}))(\forall n\in\mathbb{N})$
        $(1+\eta_n^{(1)})P_{n+1}\succeq P_n$, $(1+\eta_n^{(2)})Q_{n+1}\succeq Q_n$.
  \item[(iii)] $(\forall n\in\mathbb{N})$ $V_n\in\gP_\varsigma(\gX\times\gY)$, where
        \begin{equation*}
          V_n:=\begin{pmatrix}P_n^{-1}&-L^{*}\\-L&Q_n^{-1}\end{pmatrix}.
        \end{equation*}
\end{enumerate}
Then $\{(\vx_n,\vy_n)\}$ defined by (\ref{eq:vmpdhg}) weakly converges to a
solution of (\ref{eq:inclusion}).
\end{theorem}

\section{Learned Variable Metric (LVM)}
\label{sec:framework}
\vspace{-2mm}

\begin{figure}[t]
  \centering
  \newcommand{\lvmtile}[4]{%
  \foreach \lvmi in {0,...,3}{\foreach \lvmk in {0,...,3}{
    \pgfmathsetmacro{\lvmv}{int(8+62*abs(sin(97*\lvmi+61*\lvmk+37*#4)))}
    \fill[blue!\lvmv!white] ({#1-#3/2+\lvmi*#3/4},{#2-#3/2+\lvmk*#3/4}) rectangle ++({#3/4},{#3/4});
  }}
  \draw[black!60,line width=0.3pt] ({#1-#3/2},{#2-#3/2}) rectangle ++(#3,#3);}
\begin{tikzpicture}[
  font=\footnotesize, >=Stealth,
  box/.style={draw, rounded corners=2pt, align=center, minimum height=0.8cm,
    inner sep=3pt, fill=black!4},
  lab/.style={align=center, inner sep=1pt},
  term/.style={align=center, inner sep=1pt, font=\scriptsize},
]
  \node[box] (gen) at (0.95,0)
    {generator $G_\theta$\\[-0.5pt]{\scriptsize adapted to each}\\[-1.5pt]{\scriptsize problem instance}};
  \foreach \k in {4,3,2,1}
    \draw[black!60,line width=0.3pt,fill=white] ({2.6+0.07*\k},{-0.3+0.07*\k}) rectangle ++(0.6,0.6);
  \lvmtile{2.9}{0}{0.6}{0}
  \draw[->] (gen) -- (2.55,0);
  \node[lab] at (3.0,-0.62) {$\widetilde{P}_{\langle1\rangle},\ldots,\widetilde{P}_{\langle m\rangle}$};
  \node[term] at (3.0,-1.1) {candidate\\representative metrics};
  \draw[->] (3.55,0) -- node[above,lab] {rescaling} (5.05,0);
  \def\lvmX{5.4}\def\lvmW{3.5}\def\lvmY{-1.6}\def\lvmH{0.8}
  \foreach \j in {0,...,4} {
    \lvmtile{\lvmX+\lvmW*\j/4}{0}{0.46}{\j+1}
    \draw[black!35,densely dotted] ({\lvmX+\lvmW*\j/4},-0.26) -- ({\lvmX+\lvmW*\j/4},{\lvmY+\lvmH});
  }
  \node[lab] at (\lvmX,0.43) {$P_{\langle1\rangle}$};
  \node[term] at (\lvmX+\lvmW/2,0.43) {representative metrics};
  \node[lab] at (\lvmX+\lvmW,0.43) {$P_{\langle m\rangle}$};
  \draw[->] (\lvmX,\lvmY) -- (\lvmX+\lvmW+0.3,\lvmY) node[right,lab] {$s$};
  \draw[black!75] (\lvmX,\lvmY+\lvmH) -- (\lvmX+\lvmW/4,\lvmY);
  \foreach \j in {1,2,3}
    \draw[black!75] ({\lvmX+\lvmW*(\j-1)/4},\lvmY) -- ({\lvmX+\lvmW*\j/4},\lvmY+\lvmH) -- ({\lvmX+\lvmW*(\j+1)/4},\lvmY);
  \draw[black!75] (\lvmX+\lvmW*3/4,\lvmY) -- (\lvmX+\lvmW,\lvmY+\lvmH);
  \node[lab,below] at (\lvmX,\lvmY) {$0$};
  \node[lab,below] at (\lvmX+\lvmW,\lvmY) {$1$};
  \node[lab,anchor=east] at (\lvmX-0.08,\lvmY+0.45*\lvmH) {$\psi_j$};
  \draw[dashed,thick,red!70!black] (\lvmX+0.69*\lvmW,\lvmY) -- (\lvmX+0.69*\lvmW,\lvmY+\lvmH+0.08);
  \node[lab,below,red!70!black] at (\lvmX+0.69*\lvmW,\lvmY) {$s_n$};
  \node[box,anchor=west] (form) at (\lvmX+\lvmW+0.7,0)
    {$P_n=\Bigl(\sum_{j=1}^{m}\psi_j(s_n)P_{\langle j\rangle}^{-1}\Bigr)^{-1}$};
  \node[lab,anchor=north] at ([yshift=-3pt]form.south) {iteration $n$ of PG/PDHG};
  \draw[->] (\lvmX+\lvmW+0.35,0) -- (form.west);
\end{tikzpicture}
  \caption{Overview of LVM. Before the iterations, the generator is evaluated
  once, and its outputs are rescaled into the representative metrics. At
  iteration $n$, the metric $P_n$ combines their inverses.}
  \label{fig:overview}
  \vspace{-2mm}
\end{figure}
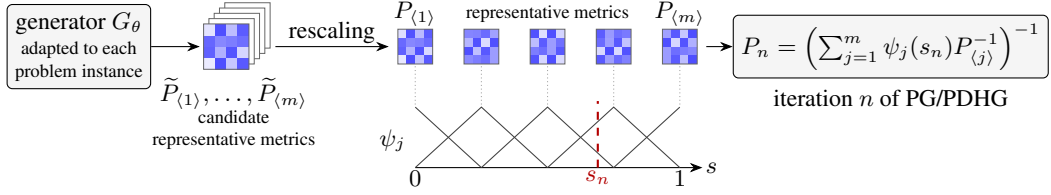

LVM constructs the metric sequence in two stages, as shown in Figure~\ref{fig:overview}.
Before the iterations, a neural network $G_\theta$ with parameters $\theta$, called the \emph{generator}, takes the data and parameters of the problem instance as input.
It outputs $m$ metrics $\widetilde{P}_{\langle1\rangle},\ldots,\widetilde{P}_{\langle m\rangle}$, whose diagonal entries are made positive by a softplus function, and they are rescaled into $P_{\langle1\rangle},\ldots,P_{\langle m\rangle}$ as described in Section~\ref{sec:convergence}.

In this work, we call $\widetilde{P}_{\langle j\rangle}$ a \emph{candidate representative metric} and $P_{\langle j\rangle}$ a \emph{representative metric}.
During the iterations, the metric of each iteration is obtained by combining the representative metrics with coefficients given by fixed functions $\psi_1,\ldots,\psi_m$, termed the \emph{basis} in the following.
Section~\ref{sec:representative} describes this construction of the metric sequence from given representative metrics, and Section~\ref{sec:convergence} specializes the representative metrics to PG and PDHG to guarantee convergence.

\subsection{Metric Sequences from Representative Metrics}
\label{sec:representative}
\vspace{-2mm}

We first describe the construction of the metric sequence from representative metrics.
Let $\gH$ be $\gX$ or $\gY$, let $m\ge2$, and let $P_{\langle1\rangle},\ldots,P_{\langle m\rangle}\in\gP_\alpha(\gH)$ be representative metrics for some $\alpha>0$.
For a constant $c>0$, we map each iteration $n\in\mathbb{N}$ to the normalized time $s_n:=n/(n+c)\in[0,1)$, and set the representative times $\sigma_j:=(j-1)/(m-1)$ for $j\in\{1,\ldots,m\}$.
The basis consists of the hat functions
\begin{equation}
\label{eq:hat}
  (\forall s\in[0,1])\quad\psi_j(s):=\max\{1-(m-1)|s-\sigma_j|,0\},
\end{equation}
and the metric of iteration $n$ is
\begin{equation}
\label{eq:lvm}
  P_n:=\biggl(\sum_{j=1}^{m}\psi_j(s_n)P_{\langle j\rangle}^{-1}\biggr)^{-1}.
\end{equation}

Since we design the representative metrics to be diagonal as described later, forming (\ref{eq:lvm}) requires only elementwise operations on two representative metrics.

The basis has several convenient properties.
At $s_n=\sigma_j$, only $\psi_j$ is nonzero and $P_n=P_{\langle j\rangle}$.
Between two consecutive representative times, only the two corresponding basis functions are nonzero, and the inverse metric changes linearly in $s_n$.
As $s_n$ increases to $1$, the $m$ basis functions cover all the iterations, without fixing the number of iterations in advance or switching to a fixed metric.

The sequence $\{P_n\}$ in (\ref{eq:lvm}) has the following basic properties that a metric sequence should satisfy to lead to algorithms with convergence guarantees.
More specifically, by (a) and (b) below, the sequence (\ref{eq:lvm}) satisfies assumptions (i) and (ii) of Theorems~\ref{thm:pg} and~\ref{thm:pdhg} for any representative metrics.
The proof of the following proposition is given in Appendix~\ref{app:basis}.

\begin{proposition}
\label{prop:sequence}
Let $\alpha>0$, let $c>0$, let $m\ge2$, and let $P_{\langle1\rangle},\ldots,P_{\langle m\rangle}\in\gP_\alpha(\gH)$.
Set $\mu:=\max_{1\le j\le m}\|P_{\langle j\rangle}\|$, and let $\{P_n\}$ be defined by (\ref{eq:lvm}).
Then the following hold.
\begin{enumerate}
  \item[(a)] $(\forall n\in\mathbb{N})$ $\mu\Id\succeq P_n\succeq\alpha\Id$.
  \item[(b)] $(\forall n\in\mathbb{N})$ ${(1+\eta_n)P_{n+1}\succeq P_n}$,
        where ${\eta_n:=\mu\|P_{n+1}^{-1}-P_n^{-1}\|}$, and
        $\{\eta_n\}\in\ell_{+}^{1}(\mathbb{N})$.
  \item[(c)] For every $n\in\mathbb{N}$, $P_n^{-1}$ is a convex combination of
        $P_{\langle1\rangle}^{-1},\ldots,P_{\langle m\rangle}^{-1}$ with the
        coefficients $\psi_1(s_n),\ldots,\psi_m(s_n)$.
\end{enumerate}
\end{proposition}

Note that Proposition~\ref{prop:sequence} relies only on the nonnegativity, the partition of unity, and the bounded variation of the basis along $\{s_n\}$, and other bases with these properties can also be used.
Thanks to the design of (\ref{eq:lvm}) that takes a convex combination of the inverse metrics, Proposition~\ref{prop:sequence}~(c) implies that assumption (iii) needs to be ensured only for the $m$ representative metrics, since (iii) is affine in the inverse metrics.

We design each metric to be diagonal, and choose its structure so that the proximity operators in the updates can still be computed efficiently.
For instance, if a function is separable over the coordinates, as the $\ell_1$ norm and the indicator of a box are, each coordinate has its own entry, and the proximity operator is computed coordinatewise (Appendix~\ref{app:structures}).
If a function acts on groups of coordinates, as the indicator of an $\ell_2$ ball does, one entry is shared within each group, and the proximity operator reduces to that of a scaled function on each group.

\subsection{Convergence Guarantees of PG/PDHG with LVM}
\label{sec:convergence}

\paragraph{Spectral rescaling of representative metrics.}
In view of Proposition~\ref{prop:sequence}, the sequence (\ref{eq:lvm}) satisfies assumptions (i) and (ii) of Theorems~\ref{thm:pg} and~\ref{thm:pdhg} for any representative metrics.
To ensure the remaining assumption (iii) for each algorithm, we introduce a spectral rescaling of the representative metrics.

Recall that the candidate representative metrics are denoted by $\widetilde{P}_{\langle j\rangle}$, and we denote by $\widetilde{Q}_{\langle j\rangle}$ those for the dual variable in PDHG.
We generate the representative metrics from these candidates by the following process with a margin $\rho\in(0,1)$:
\begin{align}
  P_{\langle j\rangle}&:=\frac{2\rho}{\beta_j}\widetilde{P}_{\langle j\rangle}
  &&\text{with}\quad\beta_j:=\|A\widetilde{P}_{\langle j\rangle}^{1/2}\|^{2}\quad\text{for PG},
  \label{eq:rescale-pg}\\
  (P_{\langle j\rangle},Q_{\langle j\rangle})&:=\frac{\rho}{\beta_j}(\widetilde{P}_{\langle j\rangle},\widetilde{Q}_{\langle j\rangle})
  &&\text{with}\quad\beta_j:=\|\widetilde{Q}_{\langle j\rangle}^{1/2}L\widetilde{P}_{\langle j\rangle}^{1/2}\|\quad\text{for PDHG},
  \label{eq:rescale-pdhg}
\end{align}
and $\{Q_n\}$ for PDHG is defined from $Q_{\langle1\rangle},\ldots,Q_{\langle m\rangle}$ by (\ref{eq:lvm}) analogously to $\{P_n\}$.
The following proposition shows that (iii) then holds for every iteration (see Appendix~\ref{app:rescaling} for the proof).

\begin{proposition}
\label{prop:rescaling}
Let $\alpha>0$ and let $\rho\in(0,1)$.
\begin{enumerate}
  \item[(a)] Suppose that $\widetilde{P}_{\langle j\rangle}\in\gP_\alpha(\gX)$ and
        $P_{\langle j\rangle}$ is given by (\ref{eq:rescale-pg}) for every
        $j\in\{1,\ldots,m\}$, and set
        ${\zeta:=(1-\rho)\min_{1\le j\le m}\|P_{\langle j\rangle}\|^{-1}}$.
        Then $\{P_n\}$ defined by (\ref{eq:lvm}) satisfies assumption (iii) of
        Theorem~\ref{thm:pg}.
  \item[(b)] Suppose that $\widetilde{P}_{\langle j\rangle}\in\gP_\alpha(\gX)$,
        $\widetilde{Q}_{\langle j\rangle}\in\gP_\alpha(\gY)$, and
        $(P_{\langle j\rangle},Q_{\langle j\rangle})$ is given by
        (\ref{eq:rescale-pdhg}) for every $j\in\{1,\ldots,m\}$, and set
        ${\varsigma:=(1-\rho)\min_{1\le j\le m}\bigl(\min\{\|P_{\langle j\rangle}\|^{-1},\|Q_{\langle j\rangle}\|^{-1}\}\bigr)}$.
        Then $\{P_n\}$ and $\{Q_n\}$ defined by (\ref{eq:lvm}) satisfy
        assumption (iii) of Theorem~\ref{thm:pdhg}.
\end{enumerate}
\end{proposition}

In our implementation, we approximate $\beta_j$ by the power iteration (Appendix~\ref{app:setup-updates}), and we verify assumption (iii) for the resulting metrics by the Lanczos method~\citep{Kuczynski1992} in Appendix~\ref{app:rescaling-impl}.

\paragraph{Convergence of PG and PDHG.}
By combining Theorems~\ref{thm:pg} and~\ref{thm:pdhg} with Propositions~\ref{prop:sequence} and~\ref{prop:rescaling}, we obtain the convergence guarantees for PG and PDHG.
We formalize these guarantees in the following theorems, which are proved in Appendix~\ref{app:rescaling}.

\begin{theorem}[PG with a quadratic term]
\label{thm:lvm-pg}
Consider the setting of Theorem~\ref{thm:pg}, i.e., $L=\Id$ and $f=\frac12\|A\cdot-\vb\|^{2}$ with $A\in\gB(\gX,\gW)$ and $\vb\in\gW$. Suppose that the solution set to
(\ref{eq:problem}) is nonempty. Let $\alpha>0$ and let
$\rho\in(0,1)$. For every $j\in\{1,\ldots,m\}$, let
$\widetilde{P}_{\langle j\rangle}\in\gP_\alpha(\gX)$, and let
$P_{\langle j\rangle}$ be given by (\ref{eq:rescale-pg}). Then $\{\vx_n\}$
defined by (\ref{eq:vmpg}) with $\{P_n\}$ defined by (\ref{eq:lvm}) weakly
converges to a solution of (\ref{eq:problem}).
\end{theorem}

\begin{theorem}[PDHG]
\label{thm:lvm-pdhg}
Suppose that the solution set to (\ref{eq:inclusion}) is nonempty. Let
$\alpha>0$ and let $\rho\in(0,1)$. For every $j\in\{1,\ldots,m\}$, let
$\widetilde{P}_{\langle j\rangle}\in\gP_\alpha(\gX)$ and
$\widetilde{Q}_{\langle j\rangle}\in\gP_\alpha(\gY)$, and let
$(P_{\langle j\rangle},Q_{\langle j\rangle})$ be given by
(\ref{eq:rescale-pdhg}). Then $\{(\vx_n,\vy_n)\}$ defined by
(\ref{eq:vmpdhg}) with $\{P_n\}$ and $\{Q_n\}$ defined by (\ref{eq:lvm})
weakly converges to a solution of (\ref{eq:inclusion}).
\end{theorem}

Importantly, these theorems do not rely on the training of the generator.
LVM therefore separates the roles of the rescaling and the training, with the former ensuring convergence and the latter providing acceleration.
In particular, the guarantees do not restrict how the candidate representative metrics are obtained, as long as they belong to $\gP_\alpha(\gX)$ or $\gP_\alpha(\gY)$ for some $\alpha>0$ (i.e., their diagonal entries are positive when $\gX$ and $\gY$ are finite dimensional).
If necessary, we can modify the output of the generator by hand after training or blend it with another preconditioner without losing the convergence guarantees.

\subsection{Training of the Metric Generator}
\label{sec:training}
\vspace{-2mm}

We implement the generator $G_\theta$ as a neural network with parameters $\theta$ that maps the data and parameters of each problem instance to the $m$ candidate representative metrics.
We train $\theta$ on a set of problem instances whose reference solutions $\vx^{\star}$ are computed in advance.
Since Theorems~\ref{thm:lvm-pg} and~\ref{thm:lvm-pdhg} hold for any $\theta$, the loss can be designed without any penalty or constraint for ensuring the convergence.
For each instance, the algorithm runs $T$ iterations with the metrics given by (\ref{eq:lvm}), and we minimize the average over the instances of
\begin{equation}
\label{eq:loss}
  \mathcal{L}(\theta):=\sum_{t\in\mathcal{T}}\omega_t\log\frac{\|\vx_t-\vx^{\star}\|^{2}}{\|\vx_0-\vx^{\star}\|^{2}},
\end{equation}
where $\vx_t$ is the (primal) iterate at iteration $t$, which depends on $\theta$, $\mathcal{T}\subset\{1,\ldots,T\}$ is a set of checkpoint iterations, and $\omega_t\ge0$ are weights.
This loss is designed to learn metrics under which the distance to the reference solution decreases in a balanced manner throughout the iterations.
Since the distance decreases by orders of magnitude along the iterations, the sum of the distances would be dominated by the early checkpoints.
Hence we balance their contributions by the logarithm, and the weights $\omega_t$ determine which checkpoints are emphasized according to their iteration $t$.
The gradient of (\ref{eq:loss}) is computed by backpropagation through the $T$ iterations and the rescaling in Section~\ref{sec:convergence}.
Since the training is performed only once in advance, LVM is particularly useful in applications that solve many problem instances of the same form, e.g., restoring images degraded in the same manner.
The detailed settings are specified in Appendices~\ref{app:setup-generator} and~\ref{app:setup-training}.

\section{Experiments}
\vspace{-2mm}
\label{sec:experiments}
\newcommand{\best}[1]{\textbf{#1}}
\newcommand{\bestsec}[1]{\underline{#1}}
We evaluate the proposed framework on two image restoration problems, deblurring and inpainting, each solved by PG and by PDHG.
PG is applied to a sparse regularization problem on wavelet coefficients, and PDHG to a total variation problem with a box constraint and a constrained $\ell_2$ data fidelity.
One generator is trained for each pair of an algorithm and an observation operator on crops of natural images, and is evaluated on standard test images of larger sizes without retraining.
We compare the number of iterations and the computation time to reach a target accuracy with fixed stepsizes, classical diagonal preconditioners, and adaptive stepsize schemes.
The detailed settings are described in Appendix~\ref{app:setup}, and an ablation study of the learned metrics is given in Appendix~\ref{app:ablation}.

\subsection{Acceleration of PG}
\vspace{-2mm}
\label{sec:exp-pg}
\begin{table*}[t]
  \centering
  \caption{Mean number of iterations of PG until $\mathcal{G}_n$ first falls below each tolerance on Set10. The best and second-best values among the methods without an inertial term are shown in bold and underlined, respectively, and the last two rows are shown for reference. The notation ``$>500$'' indicates that the criterion was not satisfied within $500$ iterations for at least one image.}
  \label{tab:pg_full_iters}
  \setlength{\tabcolsep}{3pt}
  \renewcommand{\arraystretch}{0.95}
  \resizebox{\textwidth}{!}{%
  \begin{tabular}{l c c c c c c c c}
    \toprule
    \textbf{Method} & \multicolumn{4}{c}{Deblurring} & \multicolumn{4}{c}{Inpainting} \\
    \cmidrule(lr){2-5} \cmidrule(lr){6-9}
    & $\leq10^{-3}$ & $\leq10^{-4}$ & $\leq10^{-5}$ & $\leq10^{-6}$ & $\leq10^{-3}$ & $\leq10^{-4}$ & $\leq10^{-5}$ & $\leq10^{-6}$ \\
    \midrule
    LVM (ours)
      & $\bestsec{4}$ & $\best{8}$ & $\best{23}$ & $\best{54}$ & $\best{3}$ & $\best{7}$ & $\best{34}$ & $\best{95}$ \\
    Fixed scalar~\citep{Bauschke2017}
      & $24$ & $33$ & $49$ & $\bestsec{99}$ & $47$ & $93$ & $148$ & $238$ \\
    Jacobi diag.~\citep{Combettes2014VMFBS}
      & $7$ & $36$ & $73$ & $119$ & $166$ & $308$ & $>500$ & $>500$ \\
    AdaPGM~\citep{Malitsky2024}
      & $\best{3}$ & $\bestsec{9}$ & $\bestsec{26}$ & $\best{54}$ & $\bestsec{33}$ & $\bestsec{64}$ & $\bestsec{100}$ & $\bestsec{151}$ \\
    \addlinespace[2pt]
    \multicolumn{9}{l}{\textit{With an inertial term}} \\
    FISTA~\citep{BeckTeboulle2009}
      & $3$ & $10$ & $22$ & $35$ & $24$ & $39$ & $55$ & $70$ \\
    LVM (ours) + FISTA
      & $3$ & $9$ & $16$ & $26$ & $4$ & $10$ & $22$ & $37$ \\
    \bottomrule
  \end{tabular}}
  \vspace{-3mm}
\end{table*}

\begin{figure}[t]
  \centering
  \includegraphics[width=\linewidth]{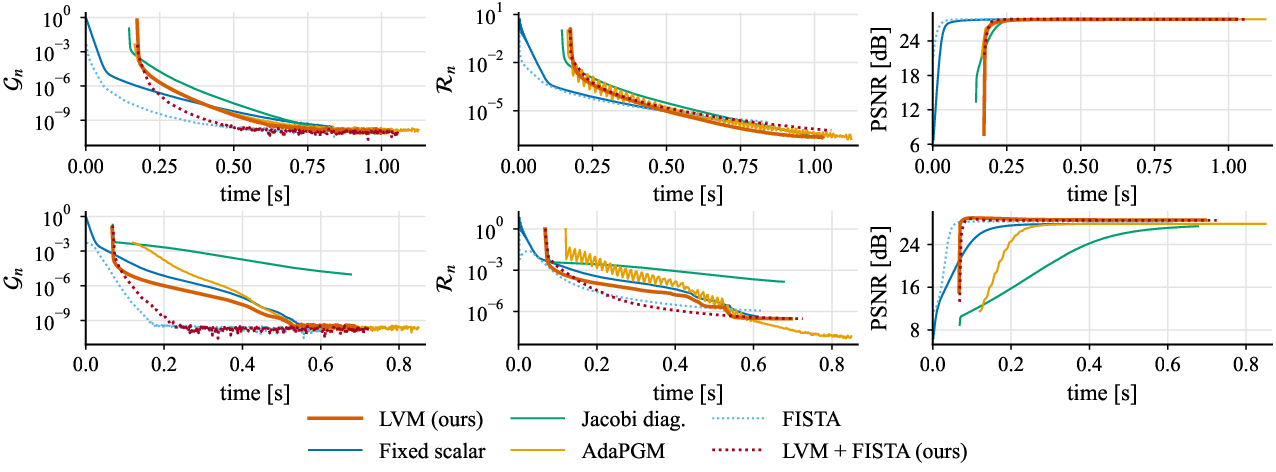}
  \caption{Convergence of PG for deblurring (top) and inpainting (bottom),
  averaged over Set10 and plotted against the computation time including the
  setup before the first iteration. The columns show the normalized
  objective gap $\mathcal{G}_n$, the relative change $\mathcal{R}_n$, and the PSNR [dB].}
  \label{fig:pg-curves}
  \vspace{-3mm}
\end{figure}

We compare LVM with four methods for PG.
Fixed scalar uses one stepsize selected by a grid search on training images and serves as the baseline without adaptation.
Jacobi diag.\ adapts the metric to each coordinate by a rule fixed in advance, using the inverse of an estimate of the diagonal of $A^{*}A$~\citep{Combettes2014VMFBS}.
AdaPGM~\citep{Malitsky2024} updates a scalar stepsize at every iteration from the curvature estimated with consecutive iterates.
FISTA~\citep{BeckTeboulle2009} accelerates the iterations by an inertial term instead of adapting the metric, and this term can be added to any choice of the metric, including LVM.
We therefore compare the metrics among the methods without an inertial term, and show FISTA and LVM + FISTA, which runs FISTA on the metrics of LVM (Appendix~\ref{app:setup-baselines}), separately.
We evaluate the convergence in terms of the normalized objective gap $\mathcal{G}_n:=(F(\vc_n)-F^{\star})/(F(\vc_0)-F^{\star})$, where $F$ is the objective function and $F^{\star}$ is the smallest value observed.
The relative change $\mathcal{R}_n:=\|\vc_n-\vc_{n-1}\|/\|\vc_n\|$ is also used as an auxiliary measure, but we note that a decrease of $\mathcal{R}_n$ does not necessarily imply a decrease of the objective value or the convergence of the algorithm in our experimental setting since it can become small even when an overly conservative stepsize or metric is chosen.

Table~\ref{tab:pg_full_iters} reports the mean number of iterations until $\mathcal{G}_n$ first falls below each tolerance.
LVM needs fewer iterations than Fixed scalar and Jacobi diag.\ in all cases, and the fewest among the methods without an inertial term for every tolerance in inpainting.
Jacobi diag.\ shares the diagonal structure with LVM but needs more iterations than Fixed scalar in inpainting, which indicates that the acceleration comes from the choice of the entries adapted to each problem instance rather than the diagonal structure alone.
We investigate the convergence against the computation time in Figure~\ref{fig:pg-curves}, where the time includes the setup before the first iteration.
For LVM, the setup is the construction of the metrics, which takes about 0.17~s in deblurring and 0.07~s in inpainting.
Since the construction is performed only once and one iteration of LVM takes only about $3\%$ longer than that of Fixed scalar, this cost is recovered as the iterations proceed, and $\mathcal{G}_n$ of LVM becomes smaller than that of Fixed scalar at around $\mathcal{G}_n=10^{-7}$ in deblurring and $10^{-3}$ in inpainting.
We also examine the iterates of each method in inpainting in Figure~\ref{fig:pg-rs-iterates} in Appendix~\ref{app:results}.
The iterate of LVM already has no visible artifacts at $n=5$, which suggests that LVM recovers coarse image structures early.

Although FISTA reaches $\mathcal{G}_n\le10^{-6}$ in fewer iterations than LVM in both tasks, LVM + FISTA attains $\mathcal{G}_n\le10^{-5}$ and $\mathcal{G}_n\le10^{-6}$ in both tasks in the fewest iterations, which suggests that the learned metrics can be combined with an inertial term.
Note that the convergence of the combination does not follow directly from Theorem~\ref{thm:pg}.

\subsection{Acceleration of PDHG}
\vspace{-1mm}
\label{sec:exp-pdhg}
\begin{table*}[t]
  \centering
  \caption{Mean number of iterations of PDHG until $\mathcal{D}_n$ first falls below each tolerance on Set10. The best and second-best values are shown in bold and underlined, respectively. The notation ``$>1000$'' indicates that the criterion was not satisfied within $1000$ iterations for at least one image.}
  \label{tab:pdhg}
  \setlength{\tabcolsep}{3pt}
  \renewcommand{\arraystretch}{0.95}
  \resizebox{\textwidth}{!}{%
  \begin{tabular}{l c c c c c c c c}
    \toprule
    \textbf{Method} & \multicolumn{4}{c}{Deblurring} & \multicolumn{4}{c}{Inpainting} \\
    \cmidrule(lr){2-5} \cmidrule(lr){6-9}
    & $\leq10^{-2}$ & $\leq10^{-3}$ & $\leq10^{-4}$ & $\leq10^{-5}$ & $\leq10^{-2}$ & $\leq10^{-3}$ & $\leq10^{-4}$ & $\leq10^{-5}$ \\
    \midrule
    LVM (ours)
      & $\best{24}$ & $\best{40}$ & $\best{62}$ & $\best{89}$ & $\best{15}$ & $\best{26}$ & $\best{39}$ & $\best{56}$ \\
    Tuned VDP~\citep{Condat2023Rev}
      & $\bestsec{47}$ & $\bestsec{77}$ & $\bestsec{115}$ & $\bestsec{172}$ & $\bestsec{20}$ & $\bestsec{34}$ & $\bestsec{56}$ & $\bestsec{121}$ \\
    P--C diag.~\citep{Pock2011Prec}
      & $>1000$ & $>1000$ & $>1000$ & $>1000$ & $22$ & $57$ & $177$ & $455$ \\
    OVDP~\citep{Naganuma2023}
      & $>1000$ & $>1000$ & $>1000$ & $>1000$ & $27$ & $66$ & $214$ & $602$ \\
    Adaptive~\citep{Goldstein2015}
      & $>1000$ & $>1000$ & $>1000$ & $>1000$ & $170$ & $330$ & $560$ & $>1000$ \\
    PDAL~\citep{Malitsky2018}
      & $>1000$ & $>1000$ & $>1000$ & $>1000$ & $32$ & $114$ & $358$ & $>1000$ \\
    \bottomrule
  \end{tabular}}
  \vspace{-3mm}
\end{table*}

\begin{figure}[t]
  \centering
  \includegraphics[width=\linewidth]{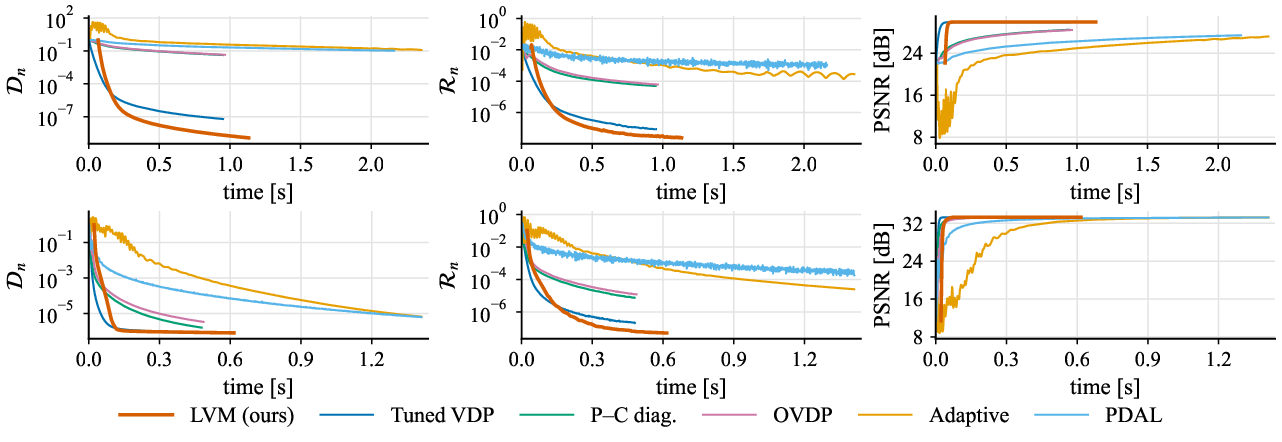}
  \caption{Convergence of PDHG for deblurring (top) and inpainting (bottom),
  averaged over Set10 and plotted against the computation time including the
  setup before the first iteration. The columns show the normalized squared
  distance $\mathcal{D}_n$, the relative change
  $\mathcal{R}_n$, and the PSNR [dB].}
  \label{fig:pdhg-curves}
  \vspace{-3mm}
\end{figure}

For PDHG, LVM is compared with five methods.
Tuned VDP assigns one stepsize to the primal variable and one to each of the two dual variables~\citep{Condat2023Rev} selected by a grid search on training images and fixed over the iterations.
P--C diag.~\citep{Pock2011Prec} and OVDP~\citep{Naganuma2023} are diagonal preconditioners given by rules fixed in advance, based on the absolute values of the entries of $L$ and on the operator norms of its blocks, respectively.
Adaptive~\citep{Goldstein2015} and PDAL~\citep{Malitsky2018} adapt scalar stepsizes along the iterations, the former by balancing the primal and dual residuals and the latter by a line search.
Since PDHG solves a saddle point problem, the objective value does not decrease monotonically, and the primal-dual gap is not directly available because the iterates may be infeasible.
We instead measure the convergence by the normalized squared distance $\mathcal{D}_n:=\|\vx_n-\vx^{\star}\|^{2}/\|\vx_0-\vx^{\star}\|^{2}$ to the reference solution $\vx^{\star}$.
The relative change $\mathcal{R}_n:=\|\vx_n-\vx_{n-1}\|/\|\vx_n\|$ of the primal variable also serves as an auxiliary measure.

Table~\ref{tab:pdhg} reports the mean number of iterations until $\mathcal{D}_n$ first falls below each tolerance.
In both tasks, LVM reaches every tolerance in the fewest iterations.
Adaptive and PDAL, which use scalar stepsizes instead of diagonal metrics, need the most iterations in inpainting, primarily because they do not exploit the block structure of $L$ in this problem.
Tuned VDP, P--C diag., and OVDP adapt to the block coordinates, but need more iterations than LVM.
We attribute this to the fact that these methods determine the adaptation by hand, either by general rules or by tuning a few stepsizes, whereas LVM represents through training a complex structure beyond the reach of such modeling.

We investigate the convergence against the computation time in Figure~\ref{fig:pdhg-curves}, where the time includes the setup before the first iteration.
For LVM, the setup is the construction of the metrics, which takes about 0.065~s in deblurring and 0.022~s in inpainting.
Since the construction is performed only once and one iteration of LVM takes about $12\%$ longer than that of Tuned VDP in deblurring and $25\%$ longer in inpainting, this cost is recovered as the iterations proceed as in the case of PG, and $\mathcal{D}_n$ of LVM becomes smaller than that of Tuned VDP at around $\mathcal{D}_n=10^{-5}$ in both tasks.

\vspace{-1mm}
\section{Conclusion}
\vspace{-2mm}
\label{sec:conclusion}
We proposed LVM, a framework that accelerates proximal splitting algorithms with metrics adapted to each problem instance, coordinate, and iteration.
The generator produces a few representative metrics once, and the metric of every iteration combines them with a fixed basis over the iterations.
With the spectral rescaling of the representative metrics, we showed that PG with a quadratic term and PDHG with LVM converge to a solution for every output of the generator when the rescaling is exact, and the assumptions of the theorems were verified numerically for the metrics used in our experiments (Appendix~\ref{app:rescaling-impl}).
In experiments on deblurring and inpainting solved by PG and PDHG, LVM leads to fewer iterations to reach a target accuracy in most tested settings than several baselines, and less computation time mainly at high accuracies.

\subsubsection*{Acknowledgments}
This work was supported by JST FOREST Grant Number JPMJFR232M, JST CREST Grant Number JPMJCR25Q5, JST ACT-X Grant Number JPMJAX26MA, and JSPS KAKENHI Grant Numbers 24K22291, 25H01296, 25K03136, and 26H02536.

\bibliography{iclr2027_conference}
\bibliographystyle{iclr2027_conference}

\clearpage
\appendix

\section{Notation and Additional Preliminaries}
\label{app:notation}

Let $M\colon\gX\to2^{\gX}$ be a set-valued operator. Its graph is
$\gra(M):=\{(\vx,\vu)\in\gX\times\gX\mid\vu\in M\vx\}$, and $M$ is monotone if
$\langle\vu_1-\vu_2\mid\vx_1-\vx_2\rangle\ge0$ for every
$(\vx_1,\vu_1)\in\gra(M)$ and $(\vx_2,\vu_2)\in\gra(M)$. If there is no monotone
operator $B$ such that $\gra(M)\subsetneq\gra(B)$, then $M$ is called maximally
monotone. The set of zeros of $M$ is $\zer(M):=\{\vx\in\gX\mid\vzero\in M\vx\}$.
The subdifferential $\partial f$ of $f\in\Gamma_0(\gX)$ is maximally monotone,
and it is related to the proximity operator by
$\prox_{f}^{U}=(\Id+U^{-1}\partial f)^{-1}$ for every $U\in\gP_\alpha(\gX)$
\citep{Bauschke2017}. Minimizing $f$ therefore amounts to finding a zero of
$\partial f$.

The product space $\gX\times\gY$ is equipped with the inner product
$\langle(\vx_1,\vy_1)\mid(\vx_2,\vy_2)\rangle:=\langle\vx_1\mid\vx_2\rangle+\langle\vy_1\mid\vy_2\rangle$. For
$L\in\gB(\gX,\gY)$, $\|L\|$ denotes the operator norm and $L^{*}$ the adjoint
operator. The proximity operator of $f^{*}$
relative to the same metric is obtained from that of $f$
\citep[Proposition 24.24 (ii)]{Bauschke2017}, which is how the dual update
(\ref{eq:vmpdhg}) is implemented.

Inversion reverses the L\"owner ordering and gives bounds on inverses, as
stated below.

\begin{proposition}[\citealp{Kato1980}, Section VI.2.6; see also
\citealp{Combettes2012VM}, Lemma 2.1]
\label{prop:inv}
Let $\gH$ be a real Hilbert space, let $\alpha>0$, let $\mu>0$, and let
$U_1$ and $U_2$ be operators in $\gS(\gH)$ such that
$\mu\Id\succeq U_1\succeq U_2\succeq\alpha\Id$. Then
$U_1^{-1},U_2^{-1}\in\gS(\gH)$,
$\alpha^{-1}\Id\succeq U_2^{-1}\succeq U_1^{-1}\succeq\mu^{-1}\Id$, and
$\|U_1^{-1}\|\le\alpha^{-1}$.
\end{proposition}

\section{Properties of the Basis and the Metric Sequence}
\label{app:basis}

This section proves Proposition~\ref{prop:sequence}. We first show the
properties of the basis (\ref{eq:hat}) used in the proof, and then derive
claims (a)--(c) from them.

\begin{lemma}
\label{lem:hat}
Let $m\ge2$ and let $c>0$. Then the following hold.
\begin{enumerate}
  \item[(a)] $(\forall s\in[0,1])$ $\psi_1(s),\ldots,\psi_m(s)\ge0$ and
        $\sum_{j=1}^{m}\psi_j(s)=1$.
  \item[(b)] $(\forall j\in\{1,\ldots,m\})$
        $\sum_{n\in\mathbb{N}}|\psi_j(s_{n+1})-\psi_j(s_n)|\le2$.
\end{enumerate}
\end{lemma}

\begin{proof}
(a) The nonnegativity follows from (\ref{eq:hat}). Let $s\in[0,1]$, and let
$k\in\{1,\ldots,m-1\}$ be such that $s\in[\sigma_k,\sigma_{k+1}]$. Since
$\sigma_{k+1}-\sigma_k=1/(m-1)$, (\ref{eq:hat}) gives $\psi_j(s)=0$ for every
$j\notin\{k,k+1\}$, and
\begin{equation*}
  \psi_k(s)+\psi_{k+1}(s)
  =\bigl(1-(m-1)(s-\sigma_k)\bigr)+\bigl(1-(m-1)(\sigma_{k+1}-s)\bigr)
  =1.
\end{equation*}

(b) Let $j\in\{1,\ldots,m\}$. Since $\{s_n\}$ is increasing in $[0,1)$, and
$\psi_j$ is nondecreasing on $[0,\sigma_j]$ and nonincreasing on
$[\sigma_j,1]$, there exists $n_0\in\mathbb{N}\cup\{+\infty\}$ such that
$\{\psi_j(s_n)\}$ is nondecreasing for $n\le n_0$ and nonincreasing for
$n\ge n_0$. Hence, by telescoping, the sum in (b) equals
$\bigl(\sup_{n\in\mathbb{N}}\psi_j(s_n)-\psi_j(s_0)\bigr)+\bigl(\sup_{n\in\mathbb{N}}\psi_j(s_n)-\lim_{n\to\infty}\psi_j(s_n)\bigr)$,
which is at most $2$.
\end{proof}

\begin{proof}[Proof of Proposition~\ref{prop:sequence}]
For every $j\in\{1,\ldots,m\}$, Proposition~\ref{prop:inv} gives
\begin{equation}
\label{eq:rep-inv}
  \alpha^{-1}\Id\succeq P_{\langle j\rangle}^{-1}\succeq\mu^{-1}\Id
  \qquad\text{and}\qquad
  \|P_{\langle j\rangle}^{-1}\|\le\alpha^{-1}.
\end{equation}

(a) Let $n\in\mathbb{N}$. By Lemma~\ref{lem:hat} (a) and (\ref{eq:rep-inv}),
$\alpha^{-1}\Id\succeq\sum_{j=1}^{m}\psi_j(s_n)P_{\langle j\rangle}^{-1}\succeq\mu^{-1}\Id$.
Hence Proposition~\ref{prop:inv} shows that $P_n$ in (\ref{eq:lvm}) is well
defined and $\mu\Id\succeq P_n\succeq\alpha\Id$.

(c) Let $n\in\mathbb{N}$. By (\ref{eq:lvm}),
$P_n^{-1}=\sum_{j=1}^{m}\psi_j(s_n)P_{\langle j\rangle}^{-1}$, and the
coefficients are nonnegative and sum to one by Lemma~\ref{lem:hat} (a).

(b) By (c) and (\ref{eq:rep-inv}), for every $n\in\mathbb{N}$,
\begin{equation*}
  \|P_{n+1}^{-1}-P_n^{-1}\|
  =\Bigl\|\sum_{j=1}^{m}\bigl(\psi_j(s_{n+1})-\psi_j(s_n)\bigr)P_{\langle j\rangle}^{-1}\Bigr\|
  \le\alpha^{-1}\sum_{j=1}^{m}|\psi_j(s_{n+1})-\psi_j(s_n)|.
\end{equation*}
Summing over $n$ and using Lemma~\ref{lem:hat} (b), we obtain
$\sum_{n\in\mathbb{N}}\eta_n\le2m\mu/\alpha<+\infty$, and hence
$\{\eta_n\}\in\ell_{+}^{1}(\mathbb{N})$. Next, let $n\in\mathbb{N}$. Then
\begin{align*}
  P_{n+1}^{-1}
  &\preceq P_n^{-1}+\|P_{n+1}^{-1}-P_n^{-1}\|\Id\\
  &\preceq P_n^{-1}+\mu\|P_{n+1}^{-1}-P_n^{-1}\|P_n^{-1}\\
  &=(1+\eta_n)P_n^{-1},
\end{align*}
where the first inequality follows from the Cauchy--Schwarz inequality, and
the second from $P_n^{-1}\succeq\mu^{-1}\Id$, which holds by (a) and
Proposition~\ref{prop:inv}. Hence, by (a) and Proposition~\ref{prop:inv},
$(1+\eta_n)P_{n+1}\succeq P_n$.
\end{proof}

\section{Convergence of Variable Metric PG}
\label{app:pg}

This section proves Theorem~\ref{thm:pg}. The proof applies the weak
convergence result of \citet{Combettes2012VM} for sequences that are
quasi-Fej\'er monotone relative to variable metrics. We show that the
sequence generated by (\ref{eq:vmpg}) is such a sequence with respect to the
solution set, where the metric at iteration $n$ is $P_n^{-1}$, and that its
weak sequential cluster points are solutions.

\paragraph{Notation.}
Throughout this section, $\{\vx_n\}$ denotes the sequence generated by
(\ref{eq:vmpg}), and we set $F:=f+g$ with $f=\frac12\|A\cdot-\vb\|^{2}$ as in
Section~\ref{sec:vm}. We write $Z$ for the solution set of (\ref{eq:problem})
with $L=\Id$, that is, the set of minimizers of $F$, which is nonempty by
assumption. The gradient of $f$ is $\nabla f=A^{*}(A\cdot-\vb)$, and hence
\begin{equation}
\label{eq:pg-grad}
  (\forall\vx\in\gX)(\forall\vy\in\gX)\qquad
  \nabla f(\vx)-\nabla f(\vy)=A^{*}A(\vx-\vy).
\end{equation}
For every $n\in\mathbb{N}$, we set $R_n:=P_n^{-1}$, which is the metric used
in (\ref{eq:vmpg}) at iteration $n$.

\paragraph{Known results.}
We recall the weak convergence of quasi-Fej\'er monotone sequences relative to
variable metrics, which is the basis of the proof. The statement below is the
``if'' part of Theorem 3.3 of \citet{Combettes2012VM}, where we set
$\phi=|\cdot|^{2}$ and $\varepsilon_n\equiv0$ in Definition 3.1 (i) there so
that the definition reduces to (\ref{eq:vmqf}). The pointwise convergence of
$\{W_n\}$ to an operator in $\gP_\alpha(\gX)$, which Theorem 3.3 requires,
follows from (\ref{eq:vmqf-metric}) by Lemma 2.3 (i) there.

\begin{proposition}[\citealp{Combettes2012VM}, Lemma 2.3 (i) and Theorem 3.3]
\label{prop:vmqf}
Let $\alpha>0$, let $\{\eta_n\}\in\ell_{+}^{1}(\mathbb{N})$, and let
$\{W_n\}$ be a sequence in $\gP_\alpha(\gX)$ such that
\begin{equation}
\label{eq:vmqf-metric}
  \sup_{n\in\mathbb{N}}\|W_n\|<+\infty
  \qquad\text{and}\qquad
  (\forall n\in\mathbb{N})\quad(1+\eta_n)W_n\succeq W_{n+1}.
\end{equation}
Let $C$ be a nonempty subset of $\gX$, and let $\{\vx_n\}$ be a sequence in
$\gX$ such that
\begin{equation}
\label{eq:vmqf}
  (\forall\vz\in C)(\forall n\in\mathbb{N})\qquad
  \|\vx_{n+1}-\vz\|_{W_{n+1}}^{2}\le(1+\eta_n)\|\vx_n-\vz\|_{W_n}^{2}.
\end{equation}
Then $\{\vx_n\}$ weakly converges to a point in $C$ if every weak sequential
cluster point of $\{\vx_n\}$ belongs to $C$.
\end{proposition}

We also use Proposition 3.2 (i) of \citet{Combettes2012VM}, which shows that
$\{\|\vx_n-\vz\|_{W_n}\}$ converges for every $\vz\in C$ under the
assumptions of Proposition~\ref{prop:vmqf}, and the bounds on inverses in
Proposition~\ref{prop:inv} of Appendix~\ref{app:notation}.

\paragraph{Auxiliary lemmas.}
For every $n\in\mathbb{N}$, we denote by $\gX_n$ the real Hilbert space
obtained by endowing $\gX$ with $\langle\cdot\mid\cdot\rangle_{R_n}$, which is
well defined by Proposition~\ref{prop:inv} and assumption (i). In $\gX_n$, the
gradient of $f$ is $P_n\nabla f$ and the proximity operator of $g$ is
$\prox_{g}^{R_n}$. The first lemma is the main step. It gives
(\ref{eq:vmqf}) with $W_n=R_n$ and $C=Z$, and it is where assumption (iii) of
Theorem~\ref{thm:pg} is used.

\begin{lemma}
\label{lem:pg-fejer}
Let $\gamma>0$ and let $\zeta>0$. Assume that assumptions (i)--(iii) of
Theorem~\ref{thm:pg} hold. Then, for every $\vz\in Z$ and every
$n\in\mathbb{N}$,
\begin{equation}
\label{eq:pg-fejer}
  \|\vx_{n+1}-\vz\|_{R_{n+1}}^{2}
  \le(1+\eta_n)\|\vx_n-\vz\|_{R_n}^{2}-\zeta\|\vx_n-\vx_{n+1}\|^{2}.
\end{equation}
\end{lemma}

\begin{proof}
Let $\vz\in Z$ and let $n\in\mathbb{N}$. Since $f$ is real-valued,
\citet[Corollary 27.3]{Bauschke2017} applied in $\gX_n$ with unit stepsize
gives $\vz=\prox_{g}^{R_n}(\vz-P_n\nabla f(\vz))$. Hence
\begin{align*}
  \|\vx_n-\vz\|_{R_n}^{2}-\|\vx_{n+1}-\vz\|_{R_n}^{2}
  &=\|\vx_n-\vx_{n+1}\|_{R_n}^{2}
    +2\langle\vx_{n+1}-\vz\mid\vx_n-\vx_{n+1}\rangle_{R_n}\\
  &\ge\|\vx_n-\vx_{n+1}\|_{R_n}^{2}
    +2\langle\vx_{n+1}-\vz\mid P_n(\nabla f(\vx_n)-\nabla f(\vz))\rangle_{R_n}\\
  &=\|\vx_n-\vx_{n+1}\|_{R_n}^{2}
    +2\langle\vx_{n+1}-\vz\mid\nabla f(\vx_n)-\nabla f(\vz)\rangle\\
  &=\|\vx_n-\vx_{n+1}\|_{R_n}^{2}
    +2\langle A(\vx_{n+1}-\vz)\mid A(\vx_n-\vz)\rangle\\
  &\ge\|\vx_n-\vx_{n+1}\|_{R_n}^{2}
    -\tfrac12\|A(\vx_n-\vx_{n+1})\|^{2}\\
  &\ge\zeta\|\vx_n-\vx_{n+1}\|^{2},
\end{align*}
where the first inequality follows from the firm nonexpansiveness of
$\prox_{g}^{R_n}$ in $\gX_n$ \citep[Proposition 12.28]{Bauschke2017}, applied
to $\vx_n-P_n\nabla f(\vx_n)$ and $\vz-P_n\nabla f(\vz)$, the second and third
equalities from $R_n=P_n^{-1}$ and (\ref{eq:pg-grad}), respectively, the
second inequality from $\|A(\vx_{n+1}+\vx_n-2\vz)\|^{2}\ge0$, and the third
from assumption (iii) of Theorem~\ref{thm:pg}. Hence
\begin{align*}
  \|\vx_{n+1}-\vz\|_{R_{n+1}}^{2}
  &\le(1+\eta_n)\|\vx_{n+1}-\vz\|_{R_n}^{2}\\
  &\le(1+\eta_n)\left(\|\vx_n-\vz\|_{R_n}^{2}
    -\zeta\|\vx_n-\vx_{n+1}\|^{2}\right)\\
  &\le(1+\eta_n)\|\vx_n-\vz\|_{R_n}^{2}-\zeta\|\vx_n-\vx_{n+1}\|^{2},
\end{align*}
where the first inequality follows from $(1+\eta_n)R_n\succeq R_{n+1}$, which
holds by Proposition~\ref{prop:inv} and assumptions (i) and (ii).
\end{proof}

The second lemma shows that the weak sequential cluster points of
$\{\vx_n\}$ are solutions.

\begin{lemma}
\label{lem:pg-cluster}
Let $\gamma>0$ and let $\zeta>0$. Assume that assumptions (i)--(iii) of
Theorem~\ref{thm:pg} hold. Then every weak sequential cluster point of
$\{\vx_n\}$ belongs to $Z$.
\end{lemma}

\begin{proof}
We first show that $\vx_n-\vx_{n+1}\to\vzero$. Fix $\vz\in Z$. By
Proposition~\ref{prop:inv}, assumptions (i)--(iii), and
Lemma~\ref{lem:pg-fejer}, the assumptions of Proposition~\ref{prop:vmqf} hold
with $W_n=R_n$ and $C=Z$. Hence Proposition 3.2 (i) of
\citet{Combettes2012VM} shows that $\{\|\vx_n-\vz\|_{R_n}\}$ converges. Since $\eta_n\to0$,
Lemma~\ref{lem:pg-fejer} yields
\begin{equation*}
  \zeta\|\vx_n-\vx_{n+1}\|^{2}
  \le(1+\eta_n)\|\vx_n-\vz\|_{R_n}^{2}
    -\|\vx_{n+1}-\vz\|_{R_{n+1}}^{2}
  \to0.
\end{equation*}
Now let $\vx$ be a weak sequential cluster point of $\{\vx_n\}$, and take a
subsequence $\{\vx_{k_n}\}$ such that $\vx_{k_n}\rightharpoonup\vx$. Then
$\vx_{k_n+1}\rightharpoonup\vx$ since $\vx_{k_n}-\vx_{k_n+1}\to\vzero$. For
every $n\in\mathbb{N}$ and every
$\vy\in\gX$,
\begin{align*}
  F(\vy)
  &\ge F(\vx_{n+1})
   +\langle\vx_n-P_n\nabla f(\vx_n)-\vx_{n+1}\mid\vy-\vx_{n+1}\rangle_{R_n}
   +\langle\nabla f(\vx_{n+1})\mid\vy-\vx_{n+1}\rangle\\
  &=F(\vx_{n+1})
   +\langle(R_n-A^{*}A)(\vx_n-\vx_{n+1})\mid\vy-\vx_{n+1}\rangle,
\end{align*}
where the inequality follows from \citet[Proposition 12.26]{Bauschke2017} in
$\gX_n$, applied to $\vx_n-P_n\nabla f(\vx_n)$, and the equality from
$R_n=P_n^{-1}$ and (\ref{eq:pg-grad}). Along
$\{k_n\}$, the inner product tends to $0$, since
$(R_{k_n}-A^{*}A)(\vx_{k_n}-\vx_{k_n+1})\to\vzero$ by
Proposition~\ref{prop:inv}, assumption (i), and $\vx_n-\vx_{n+1}\to\vzero$, and
$\{\vx_{k_n+1}\}$ is bounded \citep[Lemma 2.46]{Bauschke2017}. Since $F$ is
convex and lower semicontinuous, it is weakly lower semicontinuous
\citep[Theorem 9.1]{Bauschke2017}, and hence
$F(\vx)\le\liminf_{n\to\infty}F(\vx_{k_n+1})\le F(\vy)$ for every
$\vy\in\gX$, that is, $\vx\in Z$.
\end{proof}

\paragraph{Proof of the theorem.}
We are now ready to prove Theorem~\ref{thm:pg}.

\begin{proof}[Proof of Theorem~\ref{thm:pg}]
We apply Proposition~\ref{prop:vmqf} with $W_n=R_n$ and $C=Z$, which is
nonempty by assumption. In view of Proposition~\ref{prop:vmqf}, it remains to
verify the following.
\begin{enumerate}
  \item[(a)] $(\exists\alpha>0)(\forall n\in\mathbb{N})$ $R_n\in\gP_\alpha(\gX)$.
  \item[(b)] $\sup_{n\in\mathbb{N}}\|R_n\|<+\infty$ and
        $(\forall n\in\mathbb{N})$ $(1+\eta_n)R_n\succeq R_{n+1}$.
  \item[(c)] $(\forall\vz\in Z)(\forall n\in\mathbb{N})$
        $\|\vx_{n+1}-\vz\|_{R_{n+1}}^{2}\le(1+\eta_n)\|\vx_n-\vz\|_{R_n}^{2}$.
  \item[(d)] Every weak sequential cluster point of $\{\vx_n\}$ belongs to
        $Z$.
\end{enumerate}
Condition (a) follows from assumption (iii), (b) from
Proposition~\ref{prop:inv} and assumptions (i) and (ii), (c) from
Lemma~\ref{lem:pg-fejer}, and (d) is Lemma~\ref{lem:pg-cluster}. Hence
Proposition~\ref{prop:vmqf} shows that $\{\vx_n\}$ weakly converges to a point
in $Z$, which is a solution of (\ref{eq:problem}).
\end{proof}

\begin{remark}
\label{rem:pg-vmfbs}
The results of \citet{Combettes2014VMFBS} also cover (\ref{eq:vmpg}), but not
under assumption (iii) of Theorem~\ref{thm:pg}. Their Example 4.3 with $U_n=P_n$ requires
$\sup_{n\in\mathbb{N}}\|P_n\|<2/\|A\|^{2}$, which assumption (iii) does not imply for
nonscalar $P_n$. Their Theorem~4.1 applied to $\partial F$ alone with
$U_n=(P_n^{-1}-A^{*}A)^{-1}$ requires $P_n^{-1}-A^{*}A\succeq\zeta\Id$, which is
stronger than assumption (iii).
\end{remark}

\section{Convergence of Variable Metric PDHG}
\label{app:pdhg}

This section proves Theorem~\ref{thm:pdhg}. The work of
\citet{Suzuki2026VMPDS} assumes that
$\|P_n\|^{-1}\|Q_n\|^{-1}-\|L\|^{2}\ge\zeta$ for some $\zeta>0$, and uses this
assumption only to establish assumption (iii) of Theorem~\ref{thm:pdhg} for
some $\varsigma>0$ \citep[Lemma 3.2]{Suzuki2026VMPDS}. Theorem~\ref{thm:pdhg}
assumes the latter directly, so it covers every metric sequence covered by the
original statement, and the rest of the argument is unchanged. As in
\citet{Suzuki2026VMPDS}, the proof reduces the updates (\ref{eq:vmpdhg}) to
the proximal point algorithm on variable metrics of
\citet{Combettes2014VMFBS}.

\paragraph{Notation.}
Throughout this section, we set $\gZ:=\gX\times\gY$ and write $\vz:=(\vx,\vy)$
for its elements, $\vz_n:=(\vx_n,\vy_n)$ for the sequence generated by
(\ref{eq:vmpdhg}), and $Z$ for the solution set of (\ref{eq:inclusion}). The
space $\gZ$ is equipped with the product inner product of
Appendix~\ref{app:notation}. Each $V_n$ belongs to $\gS(\gZ)$ since $P_n^{-1}$
and $Q_n^{-1}$ are self-adjoint, and assumption (iii) of
Theorem~\ref{thm:pdhg} means that
$\langle\vz\mid V_n\vz\rangle\ge\varsigma\|\vz\|^{2}$ for every $\vz\in\gZ$.
We define the set-valued operator $M\colon\gZ\to2^{\gZ}$ by
\begin{equation}
\label{eq:pdhg-A}
  M\vz:=\begin{pmatrix}\partial f(\vx)+L^{*}\vy\\-L\vx+\partial g^{*}(\vy)\end{pmatrix},
\end{equation}
so that $Z=\zer(M)$.

\paragraph{Known results.}
We recall the proximal point algorithm on variable metrics, which is the basis
of the proof. The statement below is obtained from Theorem 4.1 (i) of
\citet{Combettes2014VMFBS} by setting $B=0$, $a_n\equiv b_n\equiv\vzero$, and
$\lambda_n\equiv\gamma_n\equiv1$ there, which is admissible since $B=0$ is
$\beta$-cocoercive for every $\beta>0$.

\begin{proposition}[\citealp{Combettes2014VMFBS}, Theorem 4.1 (i)]
\label{prop:vmpp}
Let $M\colon\gX\to2^{\gX}$ be maximally monotone such that
$\zer(M)\neq\emptyset$, let $\alpha>0$, let $\{\eta_n\}\in\ell_{+}^{1}(\mathbb{N})$,
and let $\{U_n\}$ be a sequence in $\gP_\alpha(\gX)$ such that
\begin{equation}
\label{eq:vmpp-metric}
  \sup_{n\in\mathbb{N}}\|U_n\|<+\infty
  \qquad\text{and}\qquad
  (\forall n\in\mathbb{N})\quad(1+\eta_n)U_{n+1}\succeq U_n.
\end{equation}
Let $\{\vx_n\}$ be a sequence defined by
\begin{equation}
\label{eq:vmpp}
  \vx_{n+1}=(\Id+U_nM)^{-1}(\vx_n).
\end{equation}
Then $\{\vx_n\}$ weakly converges to a point in $\zer(M)$.
\end{proposition}

We also use the bounds on inverses in Proposition~\ref{prop:inv} of
Appendix~\ref{app:notation}.

\paragraph{Auxiliary lemmas.}
The first lemma collects the bounds implied by assumptions (i) and (iii).

\begin{lemma}
\label{lem:pdhg-bounds}
Let $\gamma>0$, let $\delta>0$, and let $\varsigma>0$. Assume that assumptions
(i) and (iii) of Theorem~\ref{thm:pdhg} hold, and set
$\kappa:=\max(\gamma^{-2},\delta^{-2})$ and $\mu:=\sqrt{2(\kappa+\|L\|^{2})}$.
Then the following hold for every $n\in\mathbb{N}$.
\begin{enumerate}
  \item[(a)] $P_n^{-1}\preceq\gamma^{-1}\Id$, $Q_n^{-1}\preceq\delta^{-1}\Id$,
        $\|P_n^{-1}\|\le\gamma^{-1}$, and $\|Q_n^{-1}\|\le\delta^{-1}$.
  \item[(b)] $\|V_n\|\le\mu$.
  \item[(c)] $V_n$ is invertible, $V_n^{-1}\in\gP_{1/\mu}(\gZ)$, and
        $\|V_n^{-1}\|\le\varsigma^{-1}$.
\end{enumerate}
\end{lemma}

\begin{proof}
(a) By Proposition~\ref{prop:inv}, assumption (i) gives the claims.

(b) Let $\vz=(\vx,\vy)\in\gZ$. Then
\begin{align*}
  \|V_n\vz\|^{2}
  &=\|P_n^{-1}\vx-L^{*}\vy\|^{2}+\|{-L\vx}+Q_n^{-1}\vy\|^{2}\\
  &\le2\left(\|P_n^{-1}\vx\|^{2}+\|L^{*}\vy\|^{2}\right)
    +2\left(\|L\vx\|^{2}+\|Q_n^{-1}\vy\|^{2}\right)\\
  &\le2\left\{\left(\gamma^{-2}+\|L\|^{2}\right)\|\vx\|^{2}
    +\left(\delta^{-2}+\|L\|^{2}\right)\|\vy\|^{2}\right\}\\
  &\le2\left(\kappa+\|L\|^{2}\right)\|\vz\|^{2}=\mu^{2}\|\vz\|^{2},
\end{align*}
where the first inequality uses $(a+b)^{2}\le2a^{2}+2b^{2}$ for
$a,b\in\mathbb{R}$, and the second uses (a) and $\|L^{*}\|=\|L\|$.

(c) By Proposition~\ref{prop:inv}, (b) and assumption (iii) give the claims.
\end{proof}

The second lemma builds the summable sequence in the second condition of
(\ref{eq:vmpp-metric}) from those in assumption (ii).

\begin{lemma}
\label{lem:pdhg-order}
Let $\gamma>0$, let $\delta>0$, and let $\varsigma>0$. Assume that assumptions
(i)--(iii) of Theorem~\ref{thm:pdhg} hold, and set
\begin{equation}
\label{eq:pdhg-eta}
  (\forall n\in\mathbb{N})\qquad
  \eta_n:=\max\left\{\frac{\eta_n^{(1)}}{\varsigma\gamma},
  \frac{\eta_n^{(2)}}{\varsigma\delta}\right\}.
\end{equation}
Then $\{\eta_n\}\in\ell_{+}^{1}(\mathbb{N})$ and
\begin{equation}
\label{eq:pdhg-order}
  (\forall n\in\mathbb{N})\qquad
  (1+\eta_n)V_n\succeq V_{n+1}
  \qquad\text{and}\qquad
  (1+\eta_n)V_{n+1}^{-1}\succeq V_n^{-1}.
\end{equation}
\end{lemma}

\begin{proof}
Since $0\le\eta_n\le\eta_n^{(1)}/(\varsigma\gamma)+\eta_n^{(2)}/(\varsigma\delta)$
for every $n\in\mathbb{N}$, the sequence $\{\eta_n\}$ is summable. Let
$\vz=(\vx,\vy)\in\gZ$ and $n\in\mathbb{N}$. By the definition of $V_n$,
\begin{equation}
\label{eq:pdhg-quad}
  \langle\vz\mid V_n\vz\rangle
  =\langle\vx\mid P_n^{-1}\vx\rangle-\langle\vx\mid L^{*}\vy\rangle
   -\langle\vy\mid L\vx\rangle+\langle\vy\mid Q_n^{-1}\vy\rangle,
\end{equation}
and assumption (iii) gives
$\langle\vz\mid V_n\vz\rangle\ge\varsigma\|\vx\|^{2}+\varsigma\|\vy\|^{2}$.
Hence
\begin{equation}
\label{eq:pdhg-cross}
  \langle\vx\mid L^{*}\vy\rangle+\langle\vy\mid L\vx\rangle
  \le\langle\vx\mid P_n^{-1}\vx\rangle+\langle\vy\mid Q_n^{-1}\vy\rangle
    -\varsigma\|\vx\|^{2}-\varsigma\|\vy\|^{2}.
\end{equation}
Using (\ref{eq:pdhg-quad}) for $V_n$ and for $V_{n+1}$, and then
(\ref{eq:pdhg-cross}) with $\eta_n\ge0$, we obtain
\begin{align*}
  &(1+\eta_n)\langle\vz\mid V_n\vz\rangle-\langle\vz\mid V_{n+1}\vz\rangle\\
  &\quad=(1+\eta_n)\langle\vx\mid P_n^{-1}\vx\rangle
    -\langle\vx\mid P_{n+1}^{-1}\vx\rangle
    +(1+\eta_n)\langle\vy\mid Q_n^{-1}\vy\rangle
    -\langle\vy\mid Q_{n+1}^{-1}\vy\rangle\\
  &\qquad\quad-\eta_n\left(\langle\vx\mid L^{*}\vy\rangle
    +\langle\vy\mid L\vx\rangle\right)\\
  &\quad\ge\langle\vx\mid P_n^{-1}\vx\rangle-\langle\vx\mid P_{n+1}^{-1}\vx\rangle
    +\langle\vy\mid Q_n^{-1}\vy\rangle-\langle\vy\mid Q_{n+1}^{-1}\vy\rangle
    +\eta_n\varsigma\left(\|\vx\|^{2}+\|\vy\|^{2}\right)\\
  &\quad\ge(1+\eta_n\varsigma\gamma)\langle\vx\mid P_n^{-1}\vx\rangle
    -\langle\vx\mid P_{n+1}^{-1}\vx\rangle
    +(1+\eta_n\varsigma\delta)\langle\vy\mid Q_n^{-1}\vy\rangle
    -\langle\vy\mid Q_{n+1}^{-1}\vy\rangle\\
  &\quad\ge(1+\eta_n^{(1)})\langle\vx\mid P_n^{-1}\vx\rangle
    -\langle\vx\mid P_{n+1}^{-1}\vx\rangle
    +(1+\eta_n^{(2)})\langle\vy\mid Q_n^{-1}\vy\rangle
    -\langle\vy\mid Q_{n+1}^{-1}\vy\rangle\\
  &\quad\ge0,
\end{align*}
where the second inequality uses
$\|\vx\|^{2}\ge\gamma\langle\vx\mid P_n^{-1}\vx\rangle$ and
$\|\vy\|^{2}\ge\delta\langle\vy\mid Q_n^{-1}\vy\rangle$ from
Lemma~\ref{lem:pdhg-bounds} (a), the third uses (\ref{eq:pdhg-eta}) together
with $\langle\vx\mid P_n^{-1}\vx\rangle\ge0$ and
$\langle\vy\mid Q_n^{-1}\vy\rangle\ge0$, and the last one uses
$P_{n+1}^{-1}\preceq(1+\eta_n^{(1)})P_n^{-1}$ and
$Q_{n+1}^{-1}\preceq(1+\eta_n^{(2)})Q_n^{-1}$, which follow from
Proposition~\ref{prop:inv} and assumptions (i) and (ii). This proves the
first relation in (\ref{eq:pdhg-order}), and the second follows from it by
Proposition~\ref{prop:inv} and assumption (iii).
\end{proof}

\paragraph{Proof of the theorem.}
We are now ready to prove Theorem~\ref{thm:pdhg}.

\begin{proof}[Proof of Theorem~\ref{thm:pdhg}]
By Lemma~\ref{lem:pdhg-bounds} (c), each $V_n$ is invertible. The operator
$M$ is maximally monotone, and the updates (\ref{eq:vmpdhg}) are equivalently
rewritten as $\vz_{n+1}=(\Id+V_n^{-1}M)^{-1}(\vz_n)$
\citep[see][]{Condat2013}. This is the iteration (\ref{eq:vmpp}) with this $M$
and $U_n=V_n^{-1}$, and $\zer(M)=Z$ is nonempty by assumption. In view of
Proposition~\ref{prop:vmpp}, it remains to verify the following.
\begin{enumerate}
  \item[(a)] $(\exists\alpha>0)(\forall n\in\mathbb{N})$
        $V_n^{-1}\in\gP_\alpha(\gZ)$.
  \item[(b)] $\sup_{n\in\mathbb{N}}\|V_n^{-1}\|<+\infty$.
  \item[(c)] $(\exists\{\eta_n\}\in\ell_{+}^{1}(\mathbb{N}))(\forall n\in\mathbb{N})$
        $(1+\eta_n)V_{n+1}^{-1}\succeq V_n^{-1}$.
\end{enumerate}
Conditions (a) and (b) follow from Lemma~\ref{lem:pdhg-bounds} (c) with
$\alpha:=1/\mu$ and $\sup_{n\in\mathbb{N}}\|V_n^{-1}\|\le\varsigma^{-1}$, and (c) follows
from Lemma~\ref{lem:pdhg-order} with $\{\eta_n\}$ defined by
(\ref{eq:pdhg-eta}). Hence Proposition~\ref{prop:vmpp} shows that
$\{\vz_n\}$ weakly converges to a point in $\zer(M)=Z$.
\end{proof}

\section{Spectral Rescaling and Convergence with LVM}
\label{app:rescaling}

This section proves Proposition~\ref{prop:rescaling} and
Theorems~\ref{thm:lvm-pg} and~\ref{thm:lvm-pdhg}. In each part of
Proposition~\ref{prop:rescaling}, we first show the condition for each
representative metric, and then extend it to every iteration by
Proposition~\ref{prop:sequence} (c). The theorems follow from
Theorems~\ref{thm:pg} and~\ref{thm:pdhg} by verifying their assumptions
(i)--(iii).

\begin{proof}[Proof of Proposition~\ref{prop:rescaling}]
(a) Let $j\in\{1,\ldots,m\}$. By (\ref{eq:rescale-pg}),
$P_{\langle j\rangle}^{1/2}=(2\rho/\beta_j)^{1/2}\widetilde{P}_{\langle j\rangle}^{1/2}$,
and hence
\begin{equation}
\label{eq:margin-pg}
  \|AP_{\langle j\rangle}^{1/2}\|^{2}=\frac{2\rho}{\beta_j}\|A\widetilde{P}_{\langle j\rangle}^{1/2}\|^{2}=2\rho.
\end{equation}
Let $\vx\in\gX$. Then
\begin{align*}
  \langle\vx\mid(P_{\langle j\rangle}^{-1}-\tfrac12A^{*}A)\vx\rangle
  &=\|P_{\langle j\rangle}^{-1/2}\vx\|^{2}
    -\tfrac12\|AP_{\langle j\rangle}^{1/2}P_{\langle j\rangle}^{-1/2}\vx\|^{2}\\
  &\ge\|P_{\langle j\rangle}^{-1/2}\vx\|^{2}
    -\tfrac12\|AP_{\langle j\rangle}^{1/2}\|^{2}\|P_{\langle j\rangle}^{-1/2}\vx\|^{2}\\
  &=(1-\rho)\|P_{\langle j\rangle}^{-1/2}\vx\|^{2}\\
  &\ge(1-\rho)\|P_{\langle j\rangle}\|^{-1}\|\vx\|^{2},
\end{align*}
where the second equality follows from (\ref{eq:margin-pg}). Hence
$P_{\langle j\rangle}^{-1}-\frac12A^{*}A\succeq\zeta\Id$.

Let $n\in\mathbb{N}$. Since
$P_{\langle j\rangle}\succeq(2\rho\alpha/\beta_j)\Id$ for every $j$,
Proposition~\ref{prop:sequence} (c) applies, and
\begin{equation*}
  P_n^{-1}-\tfrac12A^{*}A
  =\sum_{j=1}^{m}\psi_j(s_n)\bigl(P_{\langle j\rangle}^{-1}-\tfrac12A^{*}A\bigr)
  \succeq\sum_{j=1}^{m}\psi_j(s_n)\zeta\Id
  =\zeta\Id.
\end{equation*}

(b) Let $j\in\{1,\ldots,m\}$. By (\ref{eq:rescale-pdhg}),
$P_{\langle j\rangle}^{1/2}=(\rho/\beta_j)^{1/2}\widetilde{P}_{\langle j\rangle}^{1/2}$
and
$Q_{\langle j\rangle}^{1/2}=(\rho/\beta_j)^{1/2}\widetilde{Q}_{\langle j\rangle}^{1/2}$,
and hence
\begin{equation}
\label{eq:margin-pdhg}
  \|Q_{\langle j\rangle}^{1/2}LP_{\langle j\rangle}^{1/2}\|
  =\frac{\rho}{\beta_j}\|\widetilde{Q}_{\langle j\rangle}^{1/2}L\widetilde{P}_{\langle j\rangle}^{1/2}\|=\rho.
\end{equation}
Let $V_{\langle j\rangle}$ be defined as $V_n$ in Theorem~\ref{thm:pdhg} with
$P_{\langle j\rangle}$ and $Q_{\langle j\rangle}$ in place of $P_n$ and $Q_n$,
and let $\vz=(\vx,\vy)\in\gX\times\gY$. Since
$\langle\vx\mid L^{*}\vy\rangle=\langle Q_{\langle j\rangle}^{1/2}LP_{\langle j\rangle}^{1/2}P_{\langle j\rangle}^{-1/2}\vx\mid Q_{\langle j\rangle}^{-1/2}\vy\rangle$,
we have
\begin{align*}
  \langle\vz\mid V_{\langle j\rangle}\vz\rangle
  &=\|P_{\langle j\rangle}^{-1/2}\vx\|^{2}
    -2\langle Q_{\langle j\rangle}^{1/2}LP_{\langle j\rangle}^{1/2}P_{\langle j\rangle}^{-1/2}\vx\mid Q_{\langle j\rangle}^{-1/2}\vy\rangle
    +\|Q_{\langle j\rangle}^{-1/2}\vy\|^{2}\\
  &\ge\|P_{\langle j\rangle}^{-1/2}\vx\|^{2}
    -2\rho\|P_{\langle j\rangle}^{-1/2}\vx\|\|Q_{\langle j\rangle}^{-1/2}\vy\|
    +\|Q_{\langle j\rangle}^{-1/2}\vy\|^{2}\\
  &\ge(1-\rho)\bigl(\|P_{\langle j\rangle}^{-1/2}\vx\|^{2}+\|Q_{\langle j\rangle}^{-1/2}\vy\|^{2}\bigr)\\
  &\ge(1-\rho)\bigl(\|P_{\langle j\rangle}\|^{-1}\|\vx\|^{2}+\|Q_{\langle j\rangle}\|^{-1}\|\vy\|^{2}\bigr)\\
  &\ge\varsigma\|\vz\|^{2},
\end{align*}
where the first inequality follows from the Cauchy--Schwarz inequality and
(\ref{eq:margin-pdhg}), and the last from the definition of $\varsigma$ in
Proposition~\ref{prop:rescaling} (b). Hence
$V_{\langle j\rangle}\in\gP_\varsigma(\gX\times\gY)$.

Let $n\in\mathbb{N}$. Since $P_{\langle j\rangle}\succeq(\rho\alpha/\beta_j)\Id$
and $Q_{\langle j\rangle}\succeq(\rho\alpha/\beta_j)\Id$ for every $j$,
Proposition~\ref{prop:sequence} (c) applies to $\{P_n\}$ and $\{Q_n\}$ with the
same coefficients $\psi_j(s_n)$. Since these coefficients sum to one and the
off-diagonal blocks of $V_{\langle j\rangle}$ do not depend on $j$, we obtain
$V_n=\sum_{j=1}^{m}\psi_j(s_n)V_{\langle j\rangle}$, and hence
$V_n\in\gP_\varsigma(\gX\times\gY)$ by the nonnegativity of the coefficients.
\end{proof}

Theorems~\ref{thm:lvm-pg} and~\ref{thm:lvm-pdhg} state the convergence of PG
and PDHG with LVM, and follow immediately by combining
Theorems~\ref{thm:pg} and~\ref{thm:pdhg} with
Propositions~\ref{prop:sequence} and~\ref{prop:rescaling}.

\begin{proof}[Proof of Theorem~\ref{thm:lvm-pg}]
Set $\nu:=2\rho\alpha\min_{1\le j\le m}\beta_j^{-1}$. By
(\ref{eq:rescale-pg}), $P_{\langle j\rangle}\in\gP_{\nu}(\gX)$ for every
$j\in\{1,\ldots,m\}$. In view of Theorem~\ref{thm:pg}, it remains to verify
its assumptions (i)--(iii).
\begin{enumerate}
  \item[(i)] This follows from Proposition~\ref{prop:sequence} (a) with
        $\nu$ in place of $\alpha$.
  \item[(ii)] This follows from Proposition~\ref{prop:sequence} (b).
  \item[(iii)] This follows from Proposition~\ref{prop:rescaling} (a).
\end{enumerate}
Hence Theorem~\ref{thm:pg} shows that $\{\vx_n\}$ weakly converges to a
solution of (\ref{eq:problem}).
\end{proof}

\begin{proof}[Proof of Theorem~\ref{thm:lvm-pdhg}]
Set $\nu:=\rho\alpha\min_{1\le j\le m}\beta_j^{-1}$. By
(\ref{eq:rescale-pdhg}), $P_{\langle j\rangle}\in\gP_{\nu}(\gX)$ and
$Q_{\langle j\rangle}\in\gP_{\nu}(\gY)$ for every $j\in\{1,\ldots,m\}$. In
view of Theorem~\ref{thm:pdhg}, it remains to verify its assumptions
(i)--(iii).
\begin{enumerate}
  \item[(i)] This follows from Proposition~\ref{prop:sequence} (a) applied to
        $\{P_n\}$ and $\{Q_n\}$ with $\nu$ in place of $\alpha$.
  \item[(ii)] This follows from Proposition~\ref{prop:sequence} (b) applied to
        $\{P_n\}$ and $\{Q_n\}$.
  \item[(iii)] This follows from Proposition~\ref{prop:rescaling} (b).
\end{enumerate}
Hence Theorem~\ref{thm:pdhg} shows that $\{(\vx_n,\vy_n)\}$ weakly converges
to a solution of (\ref{eq:inclusion}).
\end{proof}

\section{Proximity Operators Relative to Diagonal Metrics}
\label{app:structures}

Let $P=\Diag(\vp)$ with $\vp=(p_i)_{1\le i\le N}\in(0,+\infty)^{N}$, and let
$h(\vx)=\sum_{i=1}^{N}h_i(x_i)$ with $h_i\in\Gamma_0(\R)$. The objective in
(\ref{eq:prox}) with $U=P^{-1}$ is
$\sum_{i=1}^{N}\bigl(h_i(u_i)+(x_i-u_i)^{2}/(2p_i)\bigr)$, and hence it is
minimized coordinatewise, that is,
\begin{equation}
\label{eq:prox-diag}
  (\forall\vx\in\R^{N})(\forall i\in\{1,\ldots,N\})\qquad
  \bigl(\prox_{h}^{P^{-1}}(\vx)\bigr)_i=\prox_{p_ih_i}(x_i).
\end{equation}
For $h=\lambda\|\cdot\|_1$ with $\lambda>0$, (\ref{eq:prox-diag}) is the soft
thresholding with the threshold $\lambda p_i$
\citep[Example 24.20]{Bauschke2017},
\begin{equation*}
  \bigl(\prox_{\lambda\|\cdot\|_1}^{P^{-1}}(\vx)\bigr)_i
  =\operatorname{sign}(x_i)\max\{|x_i|-\lambda p_i,0\}.
\end{equation*}
For the indicator $h$ of $[0,1]^{N}$, each $p_ih_i$ is the indicator of
$[0,1]$, and (\ref{eq:prox-diag}) is the clipping $\min\{\max\{x_i,0\},1\}$,
which does not depend on $P$.

The dual update in (\ref{eq:vmpdhg}) evaluates the proximity operator of
$g^{*}$ relative to $Q_n^{-1}$, which is obtained from that of $g$ relative to
$Q_n$ by \citet[Proposition 24.24 (ii)]{Bauschke2017} as
\begin{equation}
\label{eq:prox-conj}
  \prox_{g^{*}}^{Q_n^{-1}}=Q_n\bigl(\Id-\prox_{g}^{Q_n}\bigr)Q_n^{-1}.
\end{equation}
Hence the structure of $Q_n$ is chosen according to $g$.

\section{Details of the Experimental Setup}
\label{app:setup}

\subsection{Problems and Datasets}
\label{app:setup-problems}
Let $\vx\in\R^{N}$ be a grayscale image with pixel values in $[0,1]$, and let $\vb=\Phi\vx+\vn$ be its observation, where $\vn$ is additive white Gaussian noise with standard deviation $\sigma=0.01$.
For deblurring, $\Phi$ is a circular convolution with the $19\times19$ motion blur kernel of \citet{Levin2009}.
For inpainting, $\Phi$ masks $25\%$ of the pixels chosen uniformly at random, and the noise is added before masking.
PG solves
\begin{equation}
\label{eq:pg-problem}
  \underset{\vc\in\R^{N}}{\text{minimize}}\quad\tfrac12\|\Phi\Psi^{\top}\vc-\vb\|^{2}+\lambda\|\vc\|_{1},
\end{equation}
where $\Psi$ is the three-level orthonormal Haar transform and $\vx=\Psi^{\top}\vc$, so that (\ref{eq:pg-problem}) is (\ref{eq:problem}) with $L=\Id$ and $A=\Phi\Psi^{\top}$ in Section~\ref{sec:vm}, and $\|A^{*}A\|\le1$ holds.
We set $\lambda=0.002$ for deblurring and $\lambda=0.005$ for inpainting, which are tuned manually.
PDHG solves
\begin{equation}
\label{eq:pdhg-problem}
  \underset{\vx\in[0,1]^{N}}{\text{minimize}}\quad\|D\vx\|_{2,1}\quad\text{subject to}\quad\|\Phi\vx-\vb\|\le\varepsilon,
\end{equation}
where $D$ is the forward difference operator, $\|\cdot\|_{2,1}$ is the isotropic total variation, and $\varepsilon=0.95\sigma\sqrt{N}$.
This is (\ref{eq:problem}) with $L\vx=(D\vx,\Phi\vx)$, and $\|L\|\le3$ holds.

The generators are trained on grayscale random $128\times128$ crops of the Flickr2K dataset \citep{Lim2017}, where $\lambda$ for PG is drawn at random per instance.
The test set, referred to as Set10, consists of ten standard grayscale images, seven of size $256\times256$ and three of size $512\times512$, each with one noise realization shared by PG and PDHG.

\subsection{Metrics and Updates}
\label{app:setup-updates}
PG solves (\ref{eq:pg-problem}) with the metric $P_n=\Diag(\vp_n)$, which has
one entry for each wavelet coefficient. By (\ref{eq:prox-diag}), the update
(\ref{eq:vmpg}) reads
\begin{equation*}
  \vc_{n+1}=\operatorname{soft}_{\lambda\vp_n}\bigl(\vc_n-\vp_n\odot A^{*}(A\vc_n-\vb)\bigr),
\end{equation*}
where $\odot$ is the elementwise product and $\operatorname{soft}_{\lambda\vp_n}$
is the soft thresholding with the threshold $\lambda(\vp_n)_i$ for the $i$th
coordinate.

PDHG solves (\ref{eq:pdhg-problem}) as (\ref{eq:problem}) with
$f=\iota_{[0,1]^{N}}$ and $g(\vz_1,\vz_2)=\|\vz_1\|_{2,1}+\iota_{\gC}(\vz_2)$,
where $\gC:=\{\vz\in\R^{N}\mid\|\vz-\vb\|\le\varepsilon\}$, and its dual
variable is $\vy=(\vu,\vv)\in\R^{2N}\times\R^{N}$. The primal metric
$P_n=\Diag(\vp_n)$ has one entry for each pixel. The dual metric $Q_n$
multiplies the two components of $\vu$ at the $i$th pixel by $(\vq_n)_i$ and
multiplies $\vv$ by a scalar $r_n>0$. Then the updates (\ref{eq:vmpdhg}) read
\begin{equation*}
\left\lfloor
\begin{array}{l}
  \vx_{n+1}=\operatorname{clip}\bigl(\vx_n-\vp_n\odot(D^{*}\vu_n+\Phi^{*}\vv_n)\bigr),\\[2pt]
  \vu_{n+1}=\operatorname{proj}\bigl(\vu_n+\vq_n\odot D(2\vx_{n+1}-\vx_n)\bigr),\\[2pt]
  \vv_{n+1}=\operatorname{shrink}_{r_n\varepsilon}\bigl(\vv_n+r_n(\Phi(2\vx_{n+1}-\vx_n)-\vb)\bigr),
\end{array}
\right.
\end{equation*}
where $\operatorname{clip}$ clips each pixel to $[0,1]$, $\vq_n\odot$
multiplies both components at each pixel by the same entry,
$\operatorname{proj}$ projects the two components at each pixel onto the unit
$\ell_2$ ball, and
$\operatorname{shrink}_{t}(\vw):=(1-t/\max\{\|\vw\|,t\})\vw$. The first line
follows from (\ref{eq:prox-diag}), and the others from (\ref{eq:prox-conj})
and the proximity operators of $\|\cdot\|_{2,1}$ and $\iota_{\gC}$
\citep[see][]{Parikh2014_sig}, since $Q_n$ is a multiple of the identity on
each pixel of $\vu$ and on $\vv$.

The quantity $\beta_j$ in (\ref{eq:rescale-pg}) and (\ref{eq:rescale-pdhg})
is approximated by 50 steps for PG and 20 steps for PDHG of the power iteration
applied to
$\widetilde{P}_{\langle j\rangle}^{1/2}A^{*}A\widetilde{P}_{\langle j\rangle}^{1/2}$
and to
$\widetilde{P}_{\langle j\rangle}^{1/2}L^{*}\widetilde{Q}_{\langle j\rangle}L\widetilde{P}_{\langle j\rangle}^{1/2}$,
whose largest eigenvalues are $\beta_j$ and $\beta_j^{2}$, respectively. We
multiply the estimate of $\beta_j$ by $1.05$ and set $\rho=0.95$.

\subsection{Generator}
\label{app:setup-generator}
The generator is a convolutional network with about $10^{4}$ parameters. For
PG, it acts on the wavelet coefficients arranged as an image, and its input
consists of seven channels, namely $A^{*}\vb$, $\lambda$ as a constant map, the
level of the subband of each coefficient, and four binary maps that indicate
the type of the subband. For PDHG, its input consists of two channels, $\vx_0$
and $\log(\varepsilon/\sqrt{N})$ as a constant map. A $3\times3$ convolution
with 32 channels and a ReLU are followed by six residual blocks, each
consisting of a $3\times3$ depthwise convolution with dilation 1, 2, 4, 1, 2,
and 4 in this order, a ReLU, and a $1\times1$ convolution. For PG, a
$1\times1$ convolution outputs the $m$ candidates as $m$ channels. For PDHG,
two $1\times1$ convolutions output the $m$ candidates of $\vp_n$ and $\vq_n$
in Appendix~\ref{app:setup-updates}, and global average pooling followed by
two linear layers outputs those of $r_n$. The softplus function is applied to
each output and $10^{-6}$ is added to avoid zero entries due to underflow in
single precision, although the analysis in Section~\ref{sec:convergence} does not
require this constant.

\subsection{Training}
\label{app:setup-training}
Table~\ref{tab:training} lists the settings of training. We use $m=5$, the
Adam optimizer with the learning rate $10^{-3}$, and gradient clipping at
norm 1, and keep the parameters of the epoch with the smallest average training
loss. The reference solutions of the training instances are computed by the
solvers in Appendix~\ref{app:setup-measure}, with the maximum number of
iterations and the tolerance in Table~\ref{tab:training}. In
(\ref{eq:loss}), $10^{-12}$ is added to the denominator and inside the
logarithm, and the gradient passes through the Rayleigh quotient at the last
step of the power iteration. The last layers of the generator are initialized
so that the initial metrics equal the scalar stepsize given by the rescaling
for PG and are proportional to the stepsizes of Tuned VDP in
Section~\ref{sec:exp-pdhg} for PDHG. Each generator is trained on one NVIDIA
GeForce RTX 5090 GPU.

\begin{table}[h]
  \centering
  \setlength{\tabcolsep}{2pt}
  \caption{Settings of training for each pair of an algorithm and a task.}
  \label{tab:training}
  \small
  \begin{tabular}{l c c c c}
    \toprule
    & \multicolumn{2}{c}{PG} & \multicolumn{2}{c}{PDHG}\\
    \cmidrule(lr){2-3}\cmidrule(lr){4-5}
    & Deblurring & Inpainting & Deblurring & Inpainting\\
    \midrule
    Training instances & 1000 & 10600 & 1000 & 1072\\
    Iterations and tolerance for $\vx^{\star}$ & 5000, $10^{-8}$ & 1000, $10^{-6}$ & 10000, $10^{-8}$ & 10000, $10^{-8}$\\
    $T$ & 40 & 100 & 250 & 250\\
    $\mathcal{T}$ & $\{5,10,20,40\}$ & $\{10,25,50,100\}$ & \multicolumn{2}{c}{$\{8,15,31,63,125,250\}$}\\
    $\omega_t$ & \multicolumn{2}{c}{proportional to $t$} & \multicolumn{2}{c}{uniform}\\
    $c$ & $40/9$ & $50/9$ & $250/9$ & $250/9$\\
    Batch size & 8 & 32 & 8 & 16\\
    Epochs & 100 & 30 & 100 & 200\\
    Training time [h] & 10.5 & 4.8 & 37.8 & 27.1\\
    \bottomrule
  \end{tabular}
\end{table}

\subsection{Reference Solutions and Accuracy Measures}
\label{app:setup-measure}
A reference solution of each instance is computed by FISTA for (\ref{eq:pg-problem}) and by PDHG with scalar stepsizes for (\ref{eq:pdhg-problem}), with at most $20000$ iterations and the tolerance $10^{-10}$ on the relative change of the iterate.
For each tolerance, we record the first iteration at which each measure falls below it, and report the average over the ten images only when every image reaches the tolerance within the budget of $500$ iterations for PG and $1000$ iterations for PDHG.
All methods start from $\vc_0=\vzero$ for PG and from $\vx_0=\vb$ with zero dual variables for PDHG, and run the full budget in single precision on an NVIDIA GeForce RTX 5090 GPU.
The elapsed time up to each iteration is measured by synchronizing the GPU after every iteration and is taken as the minimum over three runs that do not compute the measures.
The computation before the first iteration, such as the construction of the metrics and the power iteration, is measured separately.
Its time is included in Figures~\ref{fig:pg-curves} and~\ref{fig:pdhg-curves}.
For LVM, this computation consists of one evaluation of the generator and the power iterations, and its average time is 0.17 and 0.066 seconds for PG and 0.065 and 0.022 seconds for PDHG in deblurring and inpainting, respectively.

\subsection{Compared Methods}
\label{app:setup-baselines}
For PG, the compared methods are ISTA with a fixed scalar stepsize selected by a grid search, the Jacobi preconditioner built from an estimate of the diagonal of $A^{*}A$, the adaptive proximal gradient method of \citet{Malitsky2024}, and FISTA with the stepsize set from $\|A^{*}A\|\le1$.
For PDHG, they are a variable-wise diagonal preconditioner whose primal stepsize and two dual stepsizes~\citep{Condat2023Rev} are selected by a grid search, the diagonal preconditioner of \citet{Pock2011Prec}, the residual balancing scheme of \citet{Goldstein2015}, the line search of \citet{Malitsky2018}, and the variable-wise diagonal preconditioners of \citet{Naganuma2023}.
The grid searches score each candidate by the loss~(\ref{eq:loss}) on eight crops of Flickr2K, with $T=500$ and $\mathcal{T}=\{62,125,250,500\}$ for PG and $T=250$ and $\mathcal{T}=\{31,62,125,250\}$ for PDHG.
The initial stepsizes of the adaptive schemes are set from the power iteration for PG and from $\|L\|\le3$ for PDHG, and the remaining parameters follow the original papers.
When the learned metrics are combined with FISTA, they are scaled by a constant so that $A^{*}A\preceq P_n^{-1}$ holds, and the generator is not retrained.

\section{Details of the Experimental Results}
\label{app:results}
This section examines the early iterations of PG in inpainting with the PSNR in Figure~\ref{fig:pg-curves} and the iterates in Figure~\ref{fig:pg-rs-iterates}.
The average PSNR of LVM exceeds 28~dB at about 0.07~s, right after the construction of the metrics, whereas Fixed scalar, Jacobi diag., and AdaPGM do not reach 28~dB within 500 iterations.
In Figure~\ref{fig:pg-rs-iterates}, the iterate of LVM at $n=2$ contains no visible missing pixels and resembles a coarse approximation of the image, while the missing pixels remain visible at $n=50$ for Fixed scalar, Jacobi diag., and AdaPGM.
This suggests that the learned metrics promote the recovery of coarse image structures in the early iterations.

\newcommand{\iterdir}{results/01_vis}
\newlength{\iterw}\setlength{\iterw}{0.15\textwidth}
\newcommand{\cimg}[1]{\includegraphics[width=\iterw]{\iterdir/#1}}
\newcommand{\rlab}[1]{\rotatebox{90}{\parbox{\iterw}{\centering\scriptsize #1}}}
\newcommand{\ilab}[1]{{\scriptsize $n=#1$}}
\newcommand{\mlab}[1]{{\scriptsize #1}}

\begin{figure}[ht]
  \centering
  \setlength{\tabcolsep}{0.5pt}
  \begin{tabular}{@{}c@{\hspace{2pt}}cccccc@{}}
    & \cimg{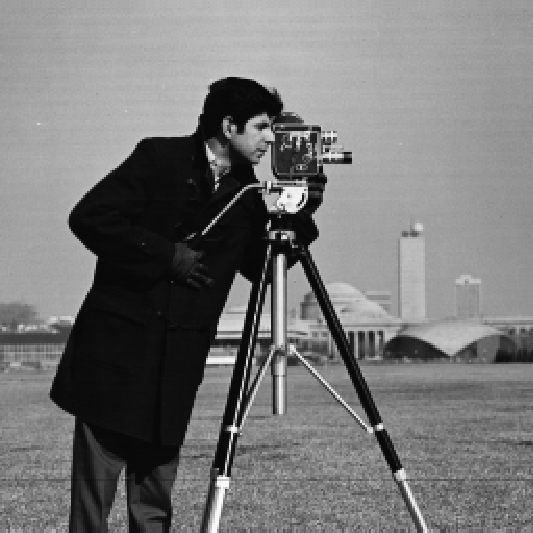} & \cimg{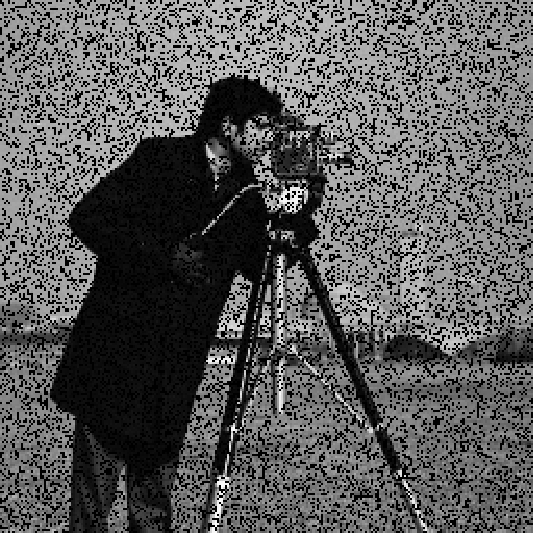} & & & & \\
    & \mlab{Ground truth} & \mlab{Observation} & & & & \\[6pt]
    & \ilab{2} & \ilab{5} & \ilab{10} & \ilab{20} & \ilab{50} & \ilab{500} \\
    \rlab{LVM (ours)}
      & \cimg{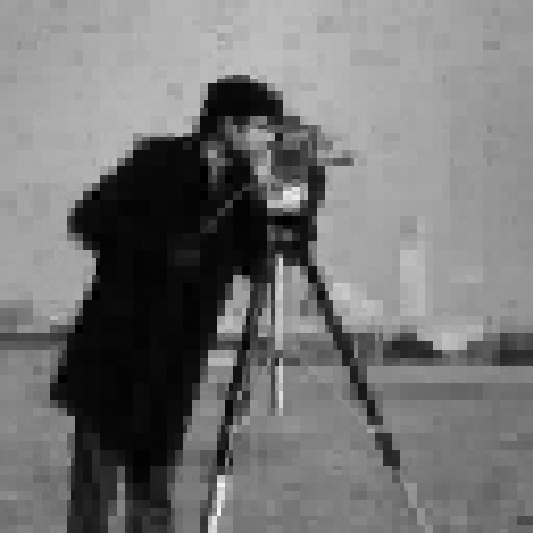} & \cimg{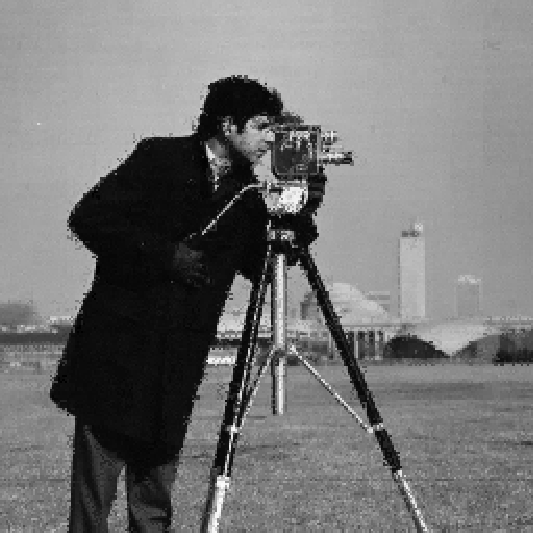} & \cimg{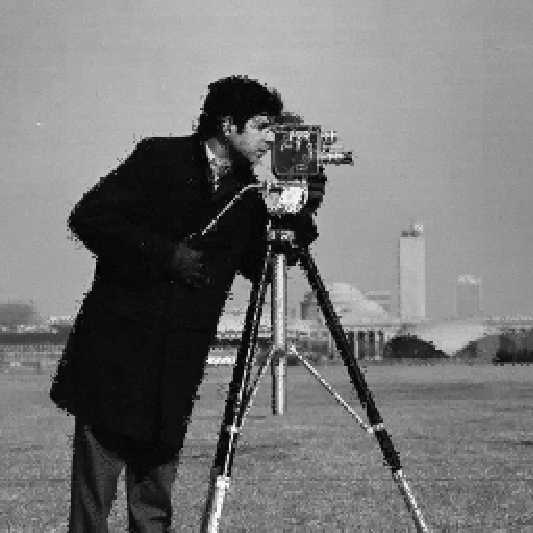} & \cimg{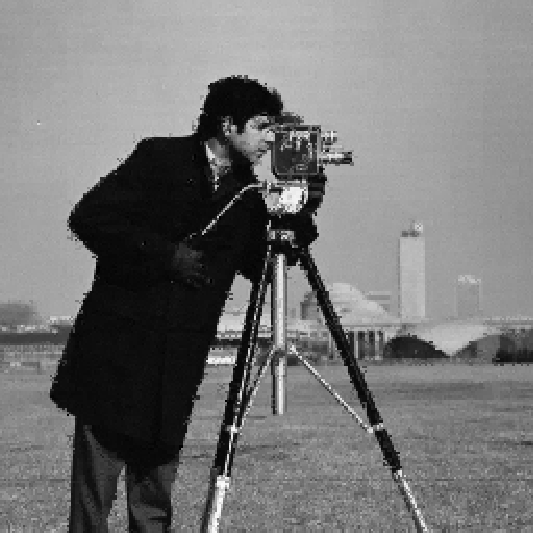} & \cimg{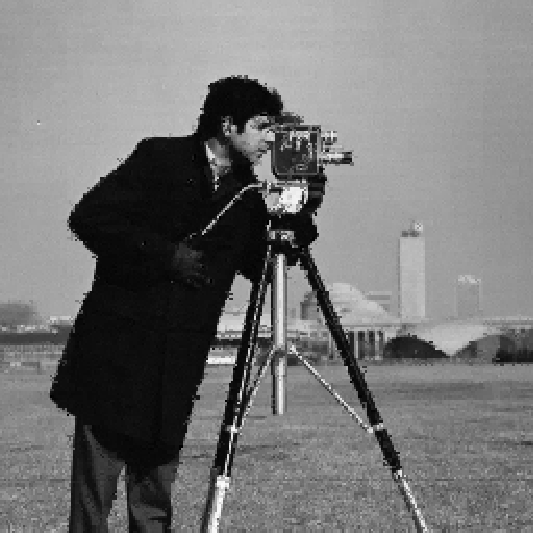} & \cimg{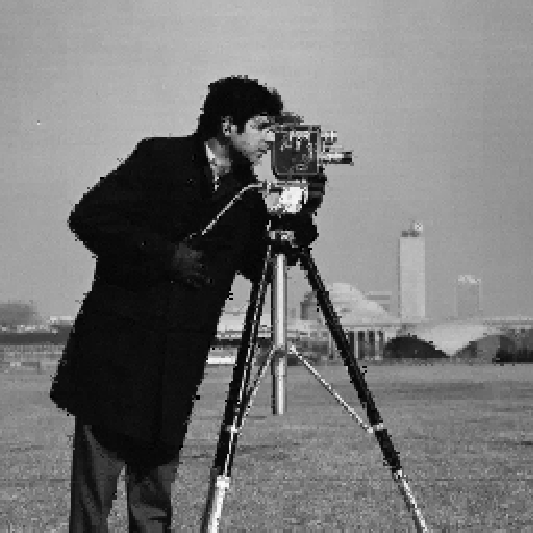} \\
    \rlab{Fixed scalar}
      & \cimg{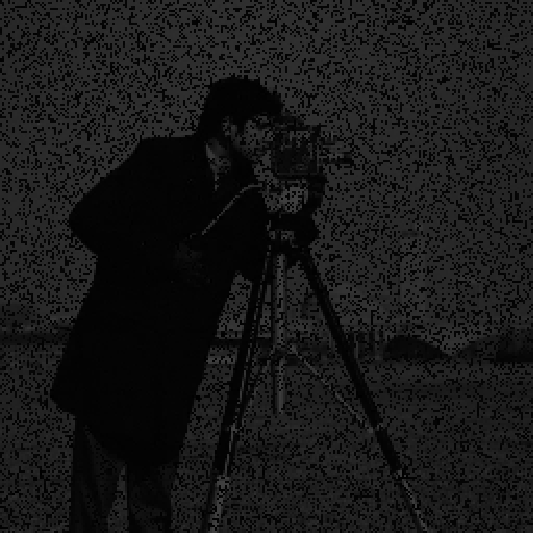} & \cimg{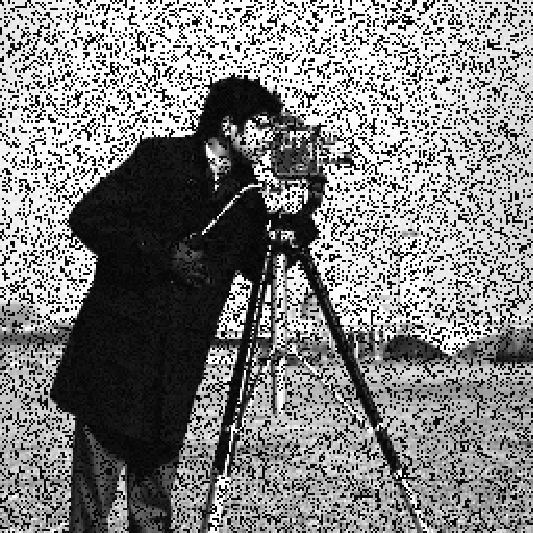} & \cimg{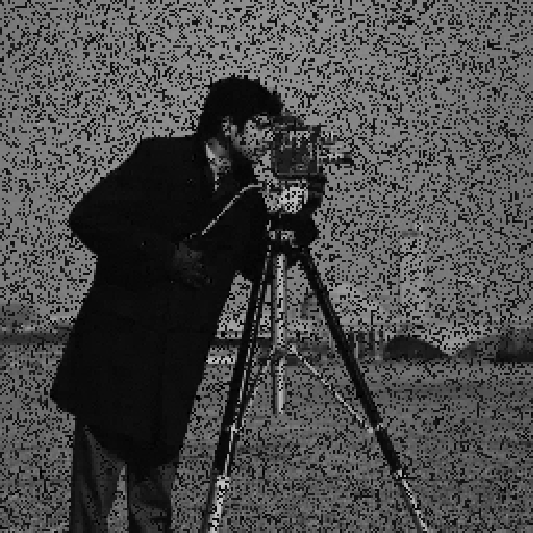} & \cimg{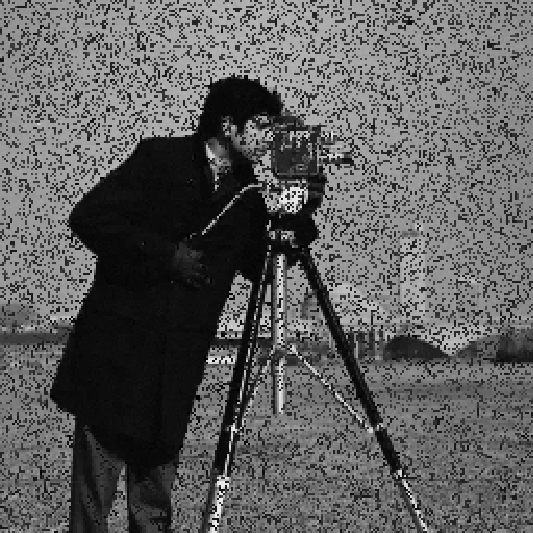} & \cimg{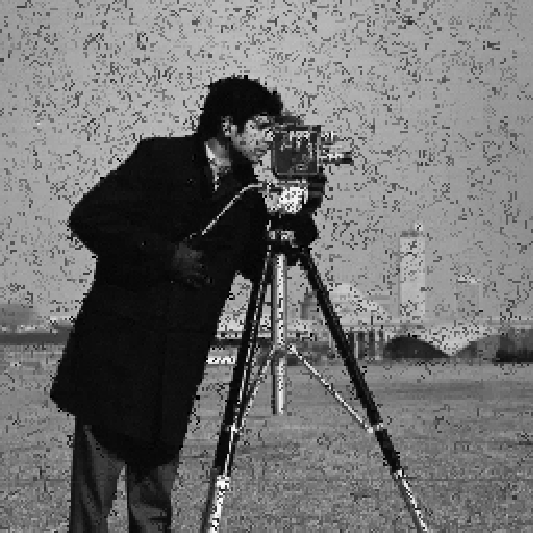} & \cimg{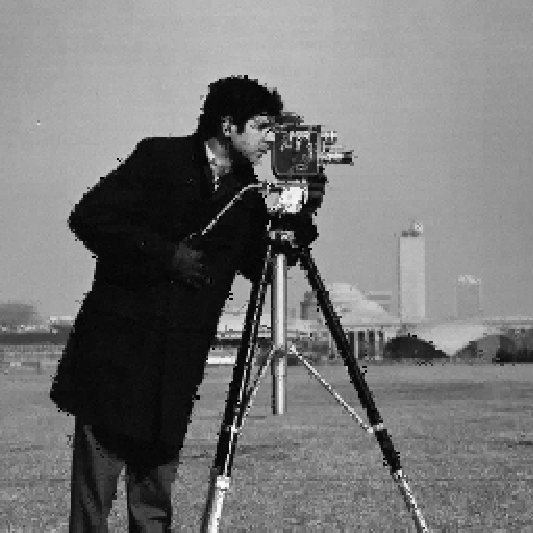} \\
    \rlab{Jacobi diag.}
      & \cimg{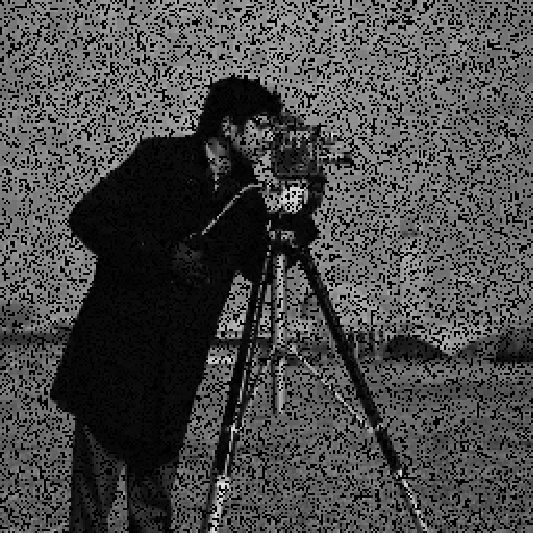} & \cimg{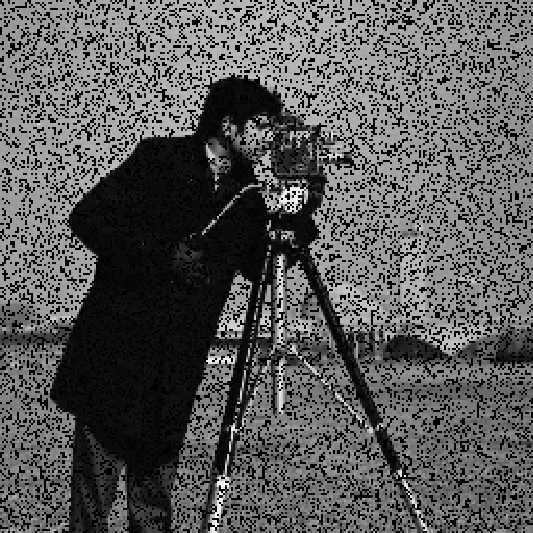} & \cimg{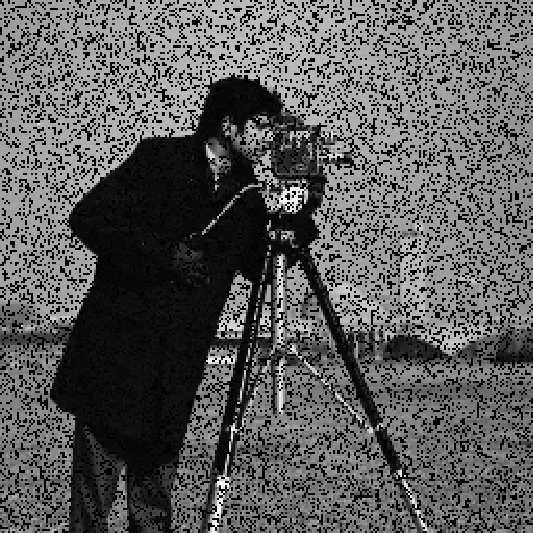} & \cimg{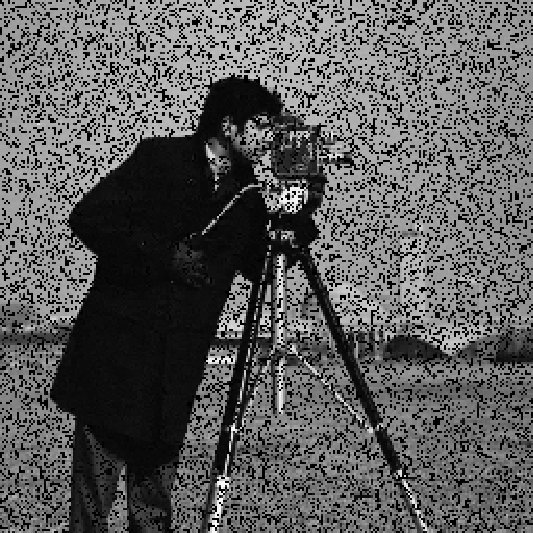} & \cimg{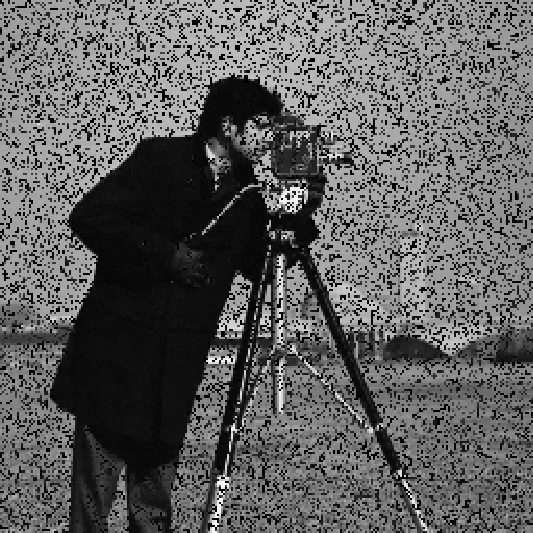} & \cimg{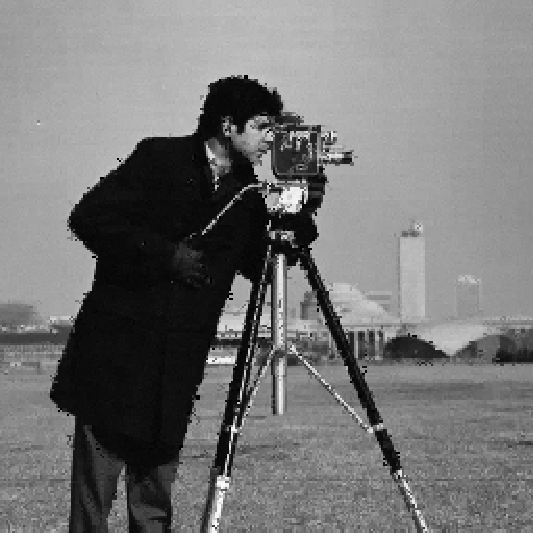} \\
    \rlab{AdaPGM}
      & \cimg{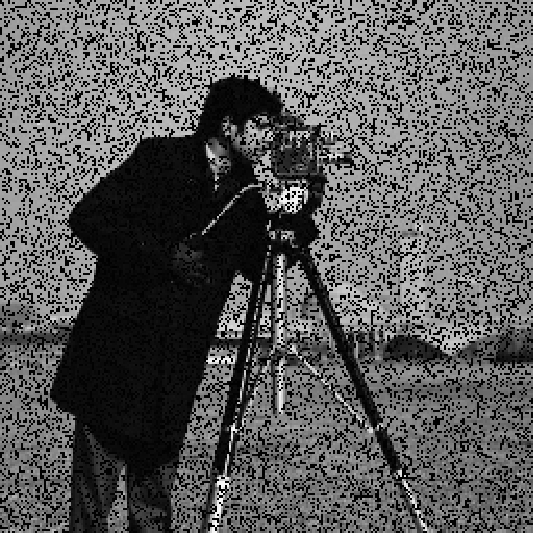} & \cimg{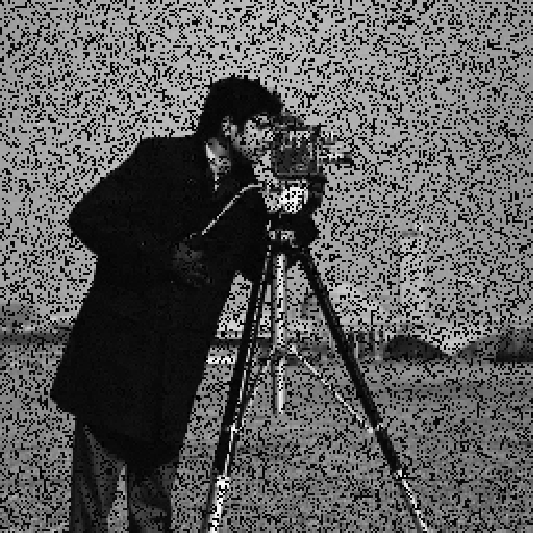} & \cimg{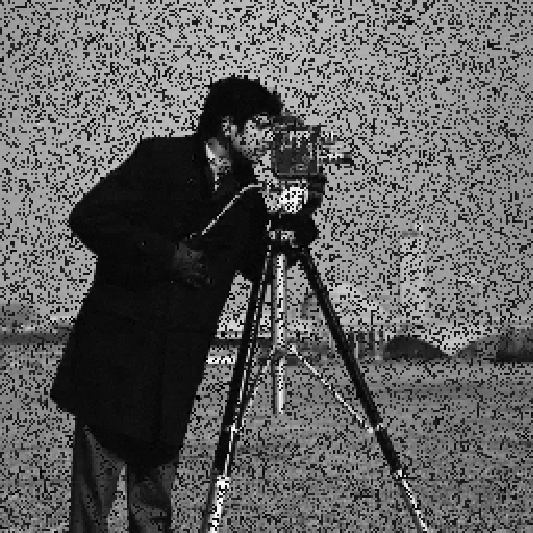} & \cimg{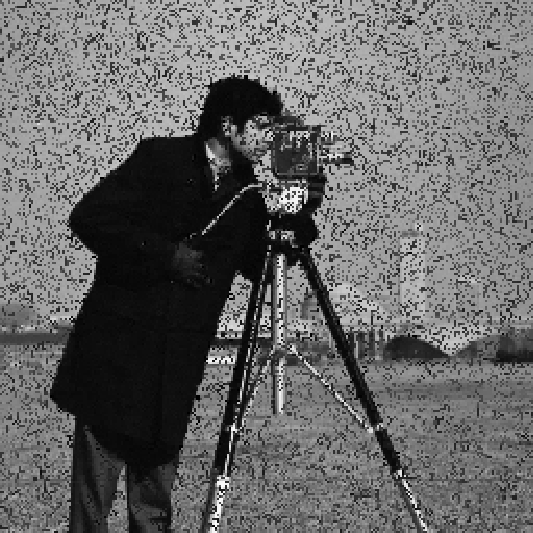} & \cimg{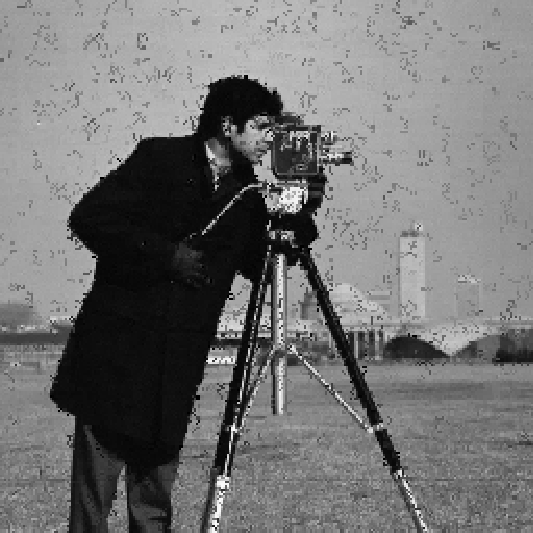} & \cimg{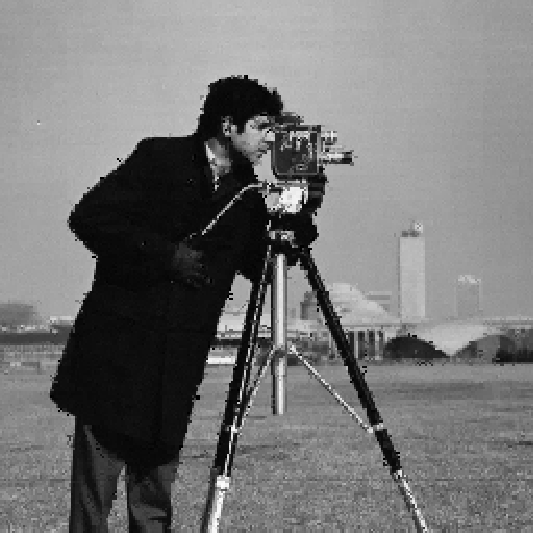} \\
    \rlab{FISTA}
      & \cimg{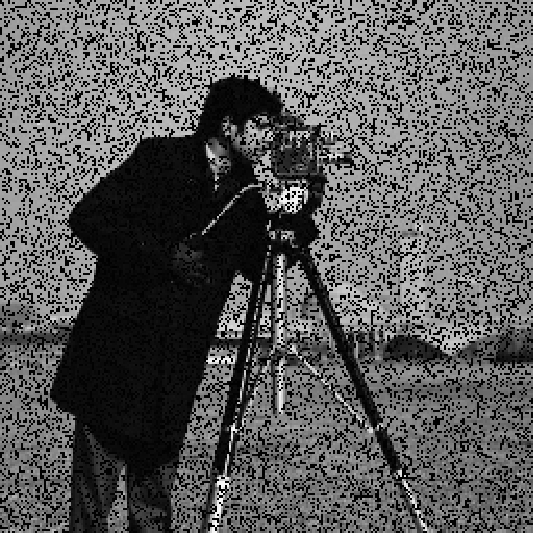} & \cimg{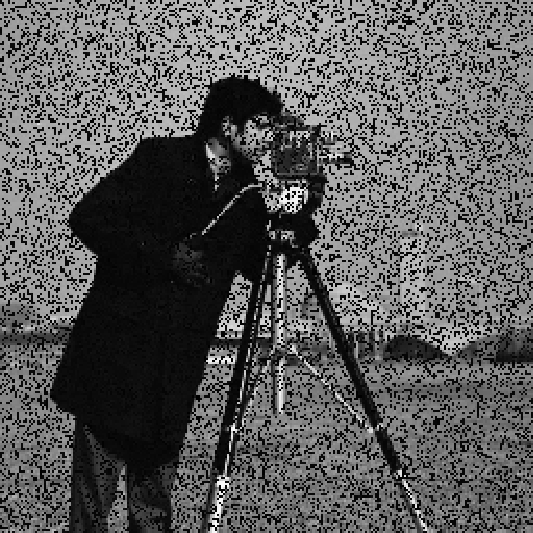} & \cimg{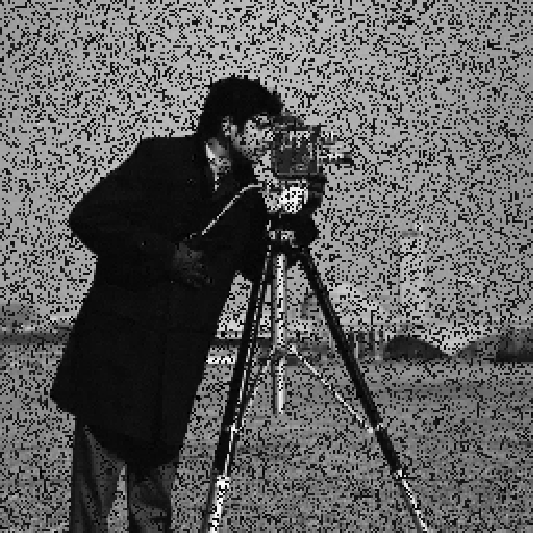} & \cimg{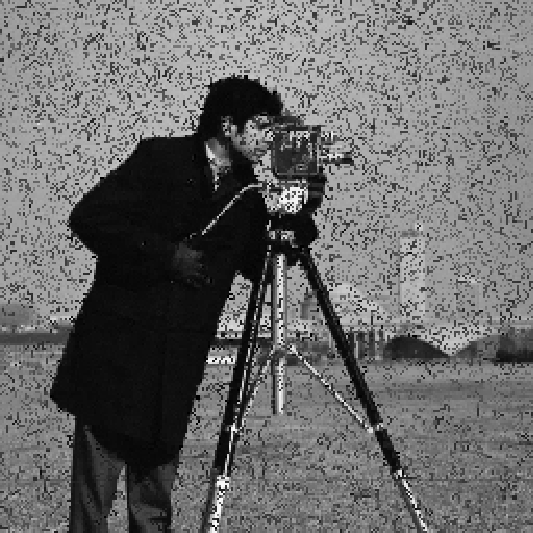} & \cimg{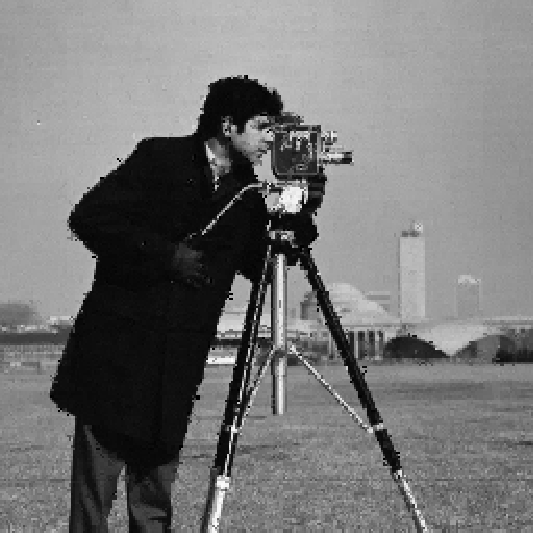} & \cimg{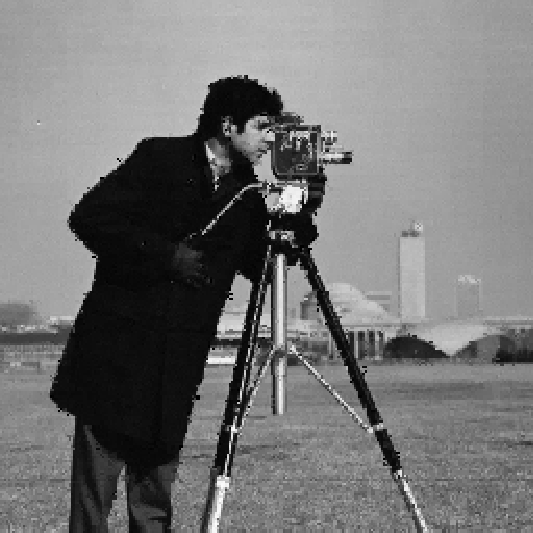} \\
  \end{tabular}
  \caption{Iterates of PG on inpainting at $n = 2, 5, 10, 20, 50$ and the
    output $n = 500$.}
  \label{fig:pg-rs-iterates}
\end{figure}

\section{Spectral Rescaling in the Implementation}
\label{app:rescaling-impl}

Proposition~\ref{prop:rescaling} assumes the exact value of $\beta_j$, whereas our implementation replaces it by
$1.05\hat\beta_j$, where $\hat\beta_j$ is obtained from the Rayleigh quotient after $K$ steps of the
power iteration, with $K = 50$ for PG and $K = 20$ for PDHG (Appendix~\ref{app:setup-updates}). Since the Rayleigh
quotient never exceeds the largest eigenvalue, $\hat\beta_j \le \beta_j$ holds, and the
representative metrics actually used satisfy $\xi_j = (\rho/1.05)(\beta_j/\hat\beta_j)$, where
$\xi_j := \frac{1}{2}\|A P_{\langle j\rangle}^{1/2}\|^2$ for PG and
$\xi_j := \|Q_{\langle j\rangle}^{1/2} L P_{\langle j\rangle}^{1/2}\|$ for PDHG. The proof of
Proposition~\ref{prop:rescaling} remains valid with $\rho$ replaced by $\max_j \xi_j$, and hence Theorems~\ref{thm:lvm-pg}
and~\ref{thm:lvm-pdhg} apply to our implementation whenever $\max_j \xi_j < 1$, which extends to every iteration
by Proposition~\ref{prop:sequence}~(c). For every problem of Set10 used in Sections~\ref{sec:exp-pg} and~\ref{sec:exp-pdhg}, we evaluated
$\xi_j$ numerically for the representative metrics actually generated, as similarly done in \citet{Pesquet2021} and \citet{Hurault2022}. 
In our evaluation, the value of $\beta_j$ is computed by the Lanczos method~\citep{Kuczynski1992} in double precision up to a relative residual of $10^{-10}$.

In Table~\ref{tab:rescaling}, the computed values of $\xi_j$ are smaller than $1$ for every setting and
every representative metric, and the largest one is $0.981$. These values suggest that the metrics used
in the experiments of the main text satisfy the assumptions of Theorems~\ref{thm:lvm-pg} and~\ref{thm:lvm-pdhg}.

\begin{table}[t]
  \centering
  \caption{Largest value of $\xi_j$ on Set10 for each representative metric $j$.
  Theorems~\ref{thm:lvm-pg} and~\ref{thm:lvm-pdhg} apply if $\xi_j < 1$ for every $j$.}
  \label{tab:rescaling}
  \small
  \begin{tabular}{l c c c c}
    \toprule
    & \multicolumn{2}{c}{PG} & \multicolumn{2}{c}{PDHG}\\
    \cmidrule(lr){2-3}\cmidrule(lr){4-5}
    & Deblurring & Inpainting & Deblurring & Inpainting\\
    \midrule
    $j=1$ & 0.937 & 0.943 & 0.942 & 0.958\\
    $j=2$ & 0.931 & 0.909 & 0.936 & 0.951\\
    $j=3$ & 0.934 & 0.952 & 0.961 & 0.981\\
    $j=4$ & 0.932 & 0.943 & 0.972 & 0.974\\
    $j=5$ & 0.931 & 0.961 & 0.949 & 0.941\\
    \bottomrule
  \end{tabular}
\end{table}

\section{Ablation Study}
\label{app:ablation}

LVM adapts its metrics to each problem instance, coordinate, and iteration.
This section examines their contributions to the acceleration by removing the variation of the metrics over the iterations or over the coordinates and the instances.

For PG, the variant without the variation over the iterations, labeled $-$~temporal variation, shares one candidate among all the representative times, and its metric still varies over the coordinates and the instances.
It follows \citet{King2024} in learning a metric fixed over the iterations.
For deblurring, it is trained with $T=200$ and $\mathcal{T}=\{5,10,20,40,100,200\}$, as is LVM (ours, matched training) in Table~\ref{tab:pg_abl}.
We do not remove the variation over the coordinates for PG, since the rescaling~(\ref{eq:rescale-pg}) maps every scalar candidate to the same fixed stepsize (Section~\ref{sec:convergence}), whose tuned version is Fixed scalar in Section~\ref{sec:exp-pg}.
For PDHG, $-$~temporal variation is defined in the same way, and the variant without the variation over the coordinates, labeled $-$~spatial variation, learns one constant for each of $\vp_n$, $\vq_n$, and $r_n$ in Appendix~\ref{app:setup-updates} at each representative time, which is shared by all the pixels and all the instances.
The two variants therefore have the properties of learned fixed metrics and learned stepsizes in Table~\ref{tab:comparison}, respectively.
The other settings are the same as those of LVM (Appendix~\ref{app:setup}).

\begin{table*}[t]
    \centering
    \caption{Ablation of the learned metric for PG on Set10. Mean number of iterations until $\mathcal{G}_n$ first falls below each tolerance. The best and second-best values are shown in bold and underlined, respectively. The notation ``$>500$'' indicates that the criterion was not satisfied within $500$ iterations for at least one image.}
    \vspace{1.5mm}
    \label{tab:pg_abl}
    \small
    \setlength{\tabcolsep}{5pt}
    \renewcommand{\arraystretch}{0.95}
    \begin{tabularx}{0.7\textwidth}{l *{4}{>{\centering\arraybackslash}X}}
      \toprule
      \textbf{Method} & $\leq10^{-3}$ & $\leq10^{-4}$ & $\leq10^{-5}$ & $\leq10^{-6}$ \\
      \midrule
      \multicolumn{5}{l}{\textit{Deblurring}} \\
      LVM (ours)
        & $\best{4}$ & $\best{8}$ & $\best{23}$ & $\bestsec{54}$ \\
      LVM (ours, matched training)
        & $\bestsec{5}$ & $\bestsec{9}$ & $\bestsec{24}$ & $\best{53}$ \\
      $-$ temporal variation
        & $7$ & $13$ & $30$ & $60$ \\
      \midrule
      \multicolumn{5}{l}{\textit{Inpainting}} \\
      LVM (ours)
        & $\best{3}$ & $\best{7}$ & $\best{34}$ & $\best{95}$ \\
      $-$ temporal variation
        & $\bestsec{13}$ & $\bestsec{66}$ & $>500$ & $>500$ \\
      \bottomrule
    \end{tabularx}
    \vspace{-3mm}
\end{table*}

\begin{table*}[!t]
  \centering
  \caption{Ablation of the learned metric for PDHG on Set10. Mean number of iterations until $\mathcal{D}_n$ first falls below each tolerance. The best and second-best values are shown in bold and underlined, respectively. The notation ``$>1000$'' indicates that the criterion was not satisfied within $1000$ iterations for at least one image.}
  \vspace{1.5mm}
  \label{tab:pdhg_abl}
  \small
  \setlength{\tabcolsep}{4pt}
  \renewcommand{\arraystretch}{0.95}
  \begin{tabularx}{0.7\textwidth}{l *{4}{>{\centering\arraybackslash}X}}
    \toprule
    \textbf{Method} & $\leq10^{-2}$ & $\leq10^{-3}$ & $\leq10^{-4}$ & $\leq10^{-5}$ \\
    \midrule
    \multicolumn{5}{l}{\textit{Deblurring}} \\
    LVM (ours)
      & $\best{24}$ & $\best{40}$ & $\best{62}$ & $\best{89}$ \\
    $-$ temporal variation
      & $29$ & $50$ & $\bestsec{73}$ & $\bestsec{101}$ \\
    $-$ spatial variation
      & $\bestsec{28}$ & $\bestsec{48}$ & $75$ & $113$ \\
    \midrule
    \multicolumn{5}{l}{\textit{Inpainting}} \\
    LVM (ours)
      & $\bestsec{15}$ & $\bestsec{26}$ & $\bestsec{39}$ & $\best{56}$ \\
    $-$ temporal variation
      & $\best{10}$ & $\best{22}$ & $\best{35}$ & $\bestsec{65}$ \\
    $-$ spatial variation
      & $16$ & $28$ & $48$ & $>1000$ \\
    \bottomrule
  \end{tabularx}
  \vspace{-3mm}
\end{table*}

At the tightest tolerance, removing either variation increases the number of iterations in every setting of Tables~\ref{tab:pg_abl} and~\ref{tab:pdhg_abl}, where LVM trained with the same settings is used for PG deblurring.
The variation over the iterations matters most in PG inpainting, where the variant without it does not reach $\mathcal{G}_n\le10^{-5}$ within 500 iterations, whereas LVM reaches $\mathcal{G}_n\le10^{-6}$ in 95 iterations.
In PDHG inpainting, this variant reaches the three loosest tolerances in fewer iterations than LVM but falls behind at the tightest one, which suggests that a single metric cannot suit both the early and the late iterations.
Without the variation over the coordinates, PDHG does not reach $\mathcal{D}_n\le10^{-5}$ within 1000 iterations in inpainting and needs 113 instead of 89 iterations to reach $\mathcal{D}_n\le10^{-5}$ in deblurring.
Since the constants of this variant are shared by all the instances, this gap reflects the adaptation both to the coordinates and to the instance.
For PG, removing the variation over the coordinates leaves only a fixed scalar stepsize, and Fixed scalar in Table~\ref{tab:pg_full_iters} needs more iterations than LVM in every setting.
In PDHG deblurring, both variants reach every tolerance in fewer iterations than Tuned VDP in Table~\ref{tab:pdhg}, and LVM, which combines their degrees of freedom, needs fewer iterations than both.
These comparisons indicate that both the variation over the iterations and the adaptation to the coordinates and the instances contribute to the acceleration.

\end{document}